\documentclass[3p,times]{elsarticle}
\usepackage{xcolor}
\usepackage{dsfont}
\usepackage{amsmath}
\usepackage{amssymb}
\usepackage{graphicx}
\usepackage{enumitem} 
\usepackage{hyperref}
\usepackage{cleveref}
\usepackage{amsthm}
\usepackage{subfigure}
\usepackage{caption}
\usepackage{algorithm}
\usepackage{algpseudocode}

\usepackage{tikz}
\usetikzlibrary{trees,arrows}
\usepackage{float} 
\usepackage{url}
\usepackage{graphicx}
\usepackage{amssymb}
\usepackage{amsmath}
\usepackage{epstopdf}
\usepackage{hyperref}
\usepackage{pdflscape}
\usepackage{xcolor}
\usepackage[font=small]{caption}
\usepackage{url}
\usepackage{dsfont}

\usepackage{xcolor}

\journal{}
\begin{document}

\begin{frontmatter}

\title{
Performance analysis of Zero
Black-Derman-Toy interest rate model 
in catastrophic events: COVID-19 case study
}

\author[label1]{Grzegorz Krzy\.zanowski}
\ead{grzegorz.krzyzanowski@pwr.edu.pl}

\author[label2]{Andr\'es Sosa}
\ead{asosa@iesta.edu.uy}

\address[label1]{Hugo Steinhaus Center,
Faculty of Pure and Applied Mathematics, Wroclaw University of Science and Technology
50-370 Wroclaw, Poland}

\address[label2]{Instituto de Estad\'istica - Facultad de Ciencias Económicas y de Administraci\'on - Universidad de la República, Uruguay} 

\begin{abstract}
In this paper we continue the research of our recent interest rate tree model called Zero Black-Derman-Toy (ZBDT) model, which includes the possibility of a jump at each step
to a practically zero interest rate. This approach allows to better match to risk of financial slowdown caused by catastrophic events. We present how to valuate a wide range of financial derivatives for such a model.  
 The  classical Black-Derman-Toy (BDT) model and novel ZBDT model are described and analogies in their calibration methodology are established.
Finally two cases of applications of the novel ZBDT model are introduced. The first of them is the hypothetical case of an S-shape term structure and decreasing volatility of yields.  The second case is an application of the ZBDT model in the  structure  of United States sovereign bonds in the current $2020$ economic slowdown caused by the Coronavirus pandemic. The objective of this study is to understand the differences presented by the valuation in both models for different derivatives.

\end{abstract}

\end{frontmatter}

\section{Introduction}
Unpredictable events in financial markets are  present in different situations and, therefore, they need to be considered. The increasing participation of agents operating in the business and the complexity of the modern financial markets has driven the use of financial derivatives. These allow investors to carry out financial risk management strategies and establish hedging strategies in case of possible adverse asset price variations.

Bonds are one of the most popular and recognized financial instruments. According to a $2019$ report of Securities Industry and Financial Markets Association, in $2018$ the global bond market reached the level of $\$102$ trillion (see \cite{sifma}).  Pricing derivatives for interest rate markets is a complex task which requires a deep understanding of the local and global economy.  In recent years, the international financial derivatives market for
interest rates is on the stage of large-scale commercialization. Such a level is reached, due to its great expansion in developed countries mostly by the over-the-counter (OTC) market. It is relevant that in the currency composition of the interest rate derivatives,  US dollar-denominated contracts are the prevailing ones. These amounted to $\$160$ trillion at end-2019, equivalent to $36\%$ of all such contracts.

Among all varieties of option-style derivatives, the most recognized are Vanilla options (i.e. European and American) and their Barrier option equivalents. Whereas a European and American options are the most classical examples of an option contract. Since European option can be used only at the time of its expiry $S$, its American analogue can be exercised at any time between $0$ and $S$. The American option is the one most listed in United States and is one of the most popular derivatives in the rest of the world (see \cite{ja3} and references therein).

The derivatives called Exotic Financial Derivatives are more complex than their Vanilla analogs. These are generally traded on the OTC market and include various variants. In the world of finance there exist a large number of exotic derivatives that by different payoff functions are manifested. In this paper we focus on the simplest exotic options, namely the Barrier Options.
Barrier options are financial derivatives where the payoff depends on whether the underlying asset’s price reaches a certain threshold during a period of time.  The main advantage of these options is that they diminish the effect of possible manipulations of the market that can happen close to the option's expiration date. They are attractive to some market participants because they are less expensive than the standard Vanilla options. In particular, Barrier options on bonds have been increasing in importance and are being widely used as the hedging instruments for risk management strategies. These complex derivatives are usually hard to manage, so they give rise to new problems in pricing and hedging with these contracts. This new principle involves the design, management, and implementation of financial instruments through which we can meet the
requirements of risk management. As an example let us mention November $1994$, when LM International (a US hedge fund) purchased $\$500$ million of “knock-in” put options from Merrill Lynch and other dealers, for Steinhardt Management, having as assets Venezuelan bonds. After buying the derivatives, the hedge funds tried to boost the prices through longing call options and the underlying bonds. 
The transaction ended in a dispute between its counterparts whether the barrier was activated and the investigation by the US Securities \&  Exchange Commission into alleged price manipulation \cite{frunza2015introduction}.

Currently, we are witnesses of a global pandemic related to the Coronavirus disease 
COVID-19. It has been an extraordinary phenomenon that  generated a deep economic crisis at all levels. Mainly, the financial markets have suffered strong turbulences that by generalized losses were expressed. Hence, economies with high volatility and uncertainty are manifested. Losses occurred in early March $2020$ became widespread. The monetary authorities reacted instantly and  in the face of the most turbulent time the United States Federal Reserve (Fed) cut interest rates to a range of $0 \%$ to $0.25 \%$. This expansive monetary policy in the United States lead to a lower cost of money. The Fed's interest rate cut was similar to that of the 2008-2009 financial crisis.

Motivated by such phenomena, the appropriate modification of the classical  Black-Derman-Toy  model (BDT) on interest rates was proposed \cite{zbdt}. This modification, the Zero Black-Derman-Toy  interest rate  model (ZBDT), departs from the BDT binary tree model \cite{bdt} in that at each time step it includes 
the possibility of a downwards jump with a small probability to a practically zero interest rate value in a zero interest rate zone. Additionally, we assume that once the ZBDT process reaches the zero interest rate zone, it remains there with high probability. In practical terms,  the initial BDT binary tree model is modified to a mixed binary-ternary tree model  to find consistent interest rates with the market term structure. To see other approaches with the same goals, see \cite{ eberlein, lewis,  martin,skew}. At the time of the COVID-19 pandemic and increasing probability of the  global diseases  appearing (see e.g. \cite{epid2, epid1}), new strategies for financial engineering have to be developed. Models widely accepted by the world of finance (such as Black-Derman-Toy and Black-Scholes) need to be modified for such scenarios like epidemics, financial crises or natural disasters. It is worth  mentioning that not always catastrophic events  have a sudden and unpredictable character. In the case of COVID-19, China became a "preview" for the rest of the world as to how serious this epidemic can be and how big an economic impact it can cause. The other countries had enough time to consider epidemic as a real danger - at least for their economies. Moreover the time between first reported cases to a significant epidemic outbreak and the necessity of a lock-down was usually weeks, giving an opportunity to preparing for extreme scenarios \cite{stat3, stat2, stat1}.

The aim of this work is to compare the performance of the classic BDT model and the novel ZBDT model for the exotic derivatives applied to an expansive monetary  policy of the United States in March 2020 due to the Coronavirus recession. 
This Section 1 is the introduction to the financial markets and economic context in 2020.
Section \ref{seccion:2} provides the class of financial derivatives to value in this article.
In  Section \ref{section:bdt-zbdt-valuation} we establish the bases of the BDT model and the way of estimating its interest rates.
Moreover we introduce the economic importance of adding to the interest rate tree a new branch with the low probability into rates following Fed's monetary policy in the ZBDT model. The way to the model calibration follows the same ideas and algorithms as in the BDT model. The methodology for valuing the financial derivatives on bonds introduced in  Section \ref{section:bdt-zbdt-valuation} is also provided.
In  Section \ref{section:comparación}, we establish a theoretical term structure (S-shape) and a decreasing volatility structure in order to compare both models through the option prices and implied volatility.
In  Section \ref{section:empirical}, we apply the model with a real data from the US term structure on a date that seems relevant to compare both models (prior to the very high volatility that happened in the financial markets).
Our concluding remarks are given in  Section \ref{section:conclusions}.

\section{Selected financial options} \label{seccion:2}

In this section, we recall the payoff functions for financial derivatives considered in the work. Path-dependent options are options whose payoff depends non-trivially on the price history of an asset.  There are two varieties of path dependent options. In this paper  we mainly focus on  a soft path dependent option, that bases its value on a single price event that occurred during the life of the option. They play an important role in OTC-markets.

In Table \ref{table:1} and Table \ref{table:2} we establish that the payoff function $f(Z_{t})$ is the gain of the option holder at the time $t$ for given underlying instrument $Z$.  In the rest of the paper $T$ is the maturity of the bond, and $S$ is the expiry time of the option, $K$ is strike, $Z_{t}$ is the value of underlying instrument at time $t$,  $\displaystyle M_{t}=\max_{\tau\in[0,t]}\left(Z_{\tau}\right)$, 
$\displaystyle m_{t}=\min_{\tau\in[0,t]}\left(Z_{\tau}\right)$  in $t\in[0,S]$ and $H^{+}$, $H^{-}$ are  upper and lower  barriers respectively.

\begin{table}[!ht]\footnotesize
\begin{center}
\begin{tabular}{|c c c|} 
\hline

&Call&Put\\ \hline
Vanilla& $\max\left(Z_{S}-K,0\right)$& $\max\left(K-Z_{S},0\right)$\\			
Knock up-and-in&$\max\left(Z_{S}-K,0\right)\mathds{1}{\{M_{S}>H^{+}\}}$& $\max\left(K-Z_{S},0\right)\mathds{1}{\{M_{S}>H^{+}\}}$\\		
Knock up-and-out& $\max\left(Z_{S}-K,0\right)\mathds{1}{\{M_{S}\leq H^{+}\}}$& $\max\left(K-Z_{S},0\right)\mathds{1}{\{M_{S}\leq H^{+}\}}$\\			
Knock down-and-in& $\max\left(Z_{S}-K,0\right)\mathds{1}{\{m_{S}<H^{-}\}}$& $\max\left(K-Z_{S},0\right)\mathds{1}{\{m_{S}<H^{-}\}}$\\				
Knock down-and-out& $\max\left(Z_{S}-K,0\right)\mathds{1}{\{m_{S}\geq H^{-}\}}$&$\max\left(K-Z_{S},0\right)\mathds{1}{\{m_{S}\geq H^{-}\}}$ \\	
Knock double-out& $\max\left(Z_{S}-K,0\right)\mathds{1}{\{M_{S}\leq H^{+},m_{S}\geq H^{-}\}}$& $\max\left(K-Z_{S},0\right)\mathds{1}{\{M_{S}\leq H^{+},m_{S}\geq H^{-}\}}$\\	
Knock double-in& $\max\left(Z_{S}-K,0\right)-\max\left(Z_{S}-K,0\right)\mathds{1}{\{M_{S}\leq H^{+},m_{S}\geq H^{-}\}}$& $\max\left(K-Z_{S},0\right)-\max\left(K-Z_{S},0\right)\mathds{1}{\{M_{S}\leq H^{+},m_{S}\geq H^{-}\}}$\\	

\hline
\end{tabular}
\caption{\label{table:1}Payoff functions for the selected European--style options.}
\end{center}
\end{table}

\begin{table}[!ht]\footnotesize
\begin{center}
\begin{tabular}{|c c c|} 
\hline

&Call&Put\\ \hline
Vanilla&$\max\left(Z_{t}-K,0\right)$& $\max\left(K-Z_{t},0\right)$\\			
Knock up-and-in&$\max\left(Z_{t}-K,0\right)\mathds{1}{\{M_{t}>H^{+}\}}$& $\max\left(K-Z_{t},0\right)\mathds{1}{\{M_{t}>H^{+}\}}$\\		
Knock up-and-out&$\max\left(Z_{t}-K,0\right)\mathds{1}{\{M_{t}\leq H^{+}\}}$& $\max\left(K-Z_{t},0\right)\mathds{1}{\{M_{t}\leq H^{+}\}}$\\			
Knock down-and-in&$\max\left(Z_{t}-K,0\right)\mathds{1}{\{m_{t}<H^{-}\}}$& $\max\left(K-Z_{t},0\right)\mathds{1}{\{m_{t}<H^{-}\}}$\\				
Knock down-and-out&$\max\left(Z_{t}-K,0\right)\mathds{1}{\{m_{t}\geq H^{-}\}}$& $\max\left(K-Z_{t},0\right)\mathds{1}{\{m_{t}\geq H^{-}\}}$\\	
Knock double-out&$\max\left(Z_{t}-K,0\right)\mathds{1}{\{M_{t}\leq H^{+},m_{t}\geq H^{-}\}}$& $\max\left(K-Z_{t},0\right)\mathds{1}{\{M_{t}\leq H^{+},m_{t}\geq H^{-}\}}$\\	Knock double-in&$\max\left(Z_{t}-K,0\right)-\max\left(Z_{t}-K,0\right)\mathds{1}{\{M_{t}\leq H^{+},m_{t}\geq H^{-}\}}$& $\max\left(K-Z_{t},0\right)-\max\left(K-Z_{t},0\right)\mathds{1}{\{M_{t}\leq H^{+},m_{t}\geq H^{-}\}}$\\	

\hline
\end{tabular}
\caption{\label{table:2}Payoff functions for the selected American--style options.}
\end{center}
\end{table}

It is important to note that as a direct conclusion from the definitions of the Barrier options we get the so called in-out parity
\begin{equation*}
 Van=Knock_{in}+Knock_{out};
\end{equation*}
where $Van$ is the price of Vanilla (plain) option, and $Knock_{in},\  Knock_{out}$ are the option prices of knock-in and knock-out of the same type and style.  This property is independent of the option pricing model.

\section{Interest Rate Models and financial derivatives valuation}\label{section:bdt-zbdt-valuation}

In this section, we briefly introduce the two models to compare: the Black-Derman-Toy (BDT) with the Zero Black-Derman-Toy (ZBDT). Moreover we present the general methodology  for valuing the options from Table \ref{table:1} and Table \ref{table:2} in both models.  The more detailed descriptions of the BDT and ZBDT models can be found in \cite{bdt} and \cite{zbdt} respectively. These  papers also establish the financial market data to be used and the way to calibrate the corresponding interest rate tree through the yield and its volatility.

\subsection{The Black-Derman-Toy  model}\label{subsection:bdt}

 The BDT model is one of the most celebrated equilibrium models for interest rates \cite{bdt}. With the model it is possible to value sovereign bonds and their financial derivatives. The model consists in a binary tree with a fixed probability of transition. One assumes that the future interest rates evolve randomly in a binomial tree  with two scenarios at each node with probability equals to $1/2$, labeled, respectively, by $d$ (for "down") and $u$ (for "up"). It is important to note that it is satisfied that  an $u$ followed by a $d$ take us to the same value as a $d$ followed by an $u$. Therefore, after $T$ periods we have $T+1$ possible states for the stochastic process modelling the interest rate. Besides, the model assumes that the evolution of the interest rate is independent to the past in each period and that the volatility only depends on time  (not on the value of the interest rate).
These hypotheses give the model the characteristics of being simple and practical for its implementation. More precisely, the model uses the current term structure in certain  yields (different maturities). In order to use the model it is also necessary to compute the yield volatility. These  values by
 a series of consecutive historical term structure can be estimated. Despite its simplicity and practicality, the BDT model has some limitations \cite{expl3, expl2, expl1}).

\begin{figure}
\centering
\subfigure[]{
\begin{tikzpicture}[scale=0.82, transform shape]
\tikzstyle{every node} = [circle, fill=gray!0,draw]
\node (11) at (0,0) {$r_{0,1}$};
\node (21) at (3,0) {$r_{1,1}$};
\node (22) at (3,2) {$r_{1,2}$};
\node (31) at (6,0) {$r_{2,1}$};
\node (32) at (6,2) {$r_{2,2}$};
\node (33) at (6,4) {$r_{2,3}$};
\tikzstyle{every node} = [circle, fill=gray!40]
\node (1121) at (1.5,0) {$\frac12$};
\node (1122) at (1.5,1) {$\frac12$};
\node (2131) at (4.5,0) {$\frac12$};
\node (2132) at (4.5,1) {$\frac12$};
\node (2232) at (4.5,2) {$\frac12$};
\node (2233) at (4.5,3) {$\frac12$};
\foreach \from/\to in {11/1121,11/1122}
\draw [-,line width=0.4mm,fill=black] (\from) -- (\to);
\foreach \from/\to in {1121/21,1122/22}
\draw [->,line width=0.4mm,fill=black] (\from) -- (\to);
\foreach \from/\to in {21/2131,21/2132,22/2232,22/2233}
\draw [-,line width=0.4mm,fill=black] (\from) -- (\to);
\foreach \from/\to in {2131/31,2132/32,2232/32,2233/33}
\draw [->,line width=0.4mm,fill=black] (\from) -- (\to);

\end{tikzpicture}
}\hspace{15mm}
\centering
\subfigure[]{
\begin{tikzpicture}[scale=0.82, transform shape]
\tikzstyle{every node} = [circle, fill=gray!0,draw]
\node (11) at (0,0) {$B_{0,1}$};
\node (21) at (3,0) {$B_{1,1}$};
\node (22) at (3,2) {$B_{1,2}$};
\node (31) at (6,0) {$B_{2,1}$};
\node (32) at (6,2) {$B_{2,2}$};
\node (33) at (6,4) {$B_{2,3}$};
\node (41) at (9,0) {$100$};
\node (42) at (9,2) {$100$};
\node (43) at (9,4) {$100$};
\node (44) at (9,6) {$100$}; 
\tikzstyle{every node} = [circle, fill=gray!40]
\node (1121) at (1.5,0) {$r_{0,1}$};
\node (1122) at (1.5,1) {$r_{0,1}$};
\node (2131) at (4.5,0) {$r_{1,1}$};
\node (2132) at (4.5,1) {$r_{1,1}$};
\node (2232) at (4.5,2) {$r_{1,2}$};
\node (2233) at (4.5,3) {$r_{1,2}$};
\node (3141) at (7.5,0) {$r_{2,1}$};
\node (3142) at (7.5,1) {$r_{2,1}$};
\node (3242) at (7.5,2) {$r_{2,2}$};
\node (3243) at (7.5,3) {$r_{2,2}$};
\node (3343) at (7.5,4) {$r_{2,3}$};
\node (3344) at (7.5,5) {$r_{2,3}$};
\foreach \from/\to in {11/1121,11/1122}
\draw [-,line width=0.4mm,fill=black] (\from) -- (\to);
\foreach \from/\to in {1121/21,1122/22}
\draw [->,line width=0.4mm,fill=black] (\from) -- (\to);
\foreach \from/\to in {21/2131,21/2132,22/2232,22/2233}
\draw [-,line width=0.4mm,fill=black] (\from) -- (\to);
\foreach \from/\to in {2131/31,2132/32,2232/32,2233/33}
\draw [->,line width=0.4mm,fill=black] (\from) -- (\to);
\foreach \from/\to in {31/3141,31/3142,32/3242,32/3243,33/3343,33/3344}
\draw [-,line width=0.4mm,fill=black] (\from) -- (\to);
\foreach \from/\to in {3141/41,3142/42,3242/42,3243/43,3343/43,3344/44}
\draw [->,line width=0.4mm,fill=black] (\from) -- (\to);
\end{tikzpicture}
}

\caption{The BDT interest rate tree (a) and the corresponding tree of zc-bond prices (b) with $T=3$ and $FV=100$.}\label{figure:bdtp}
\end{figure}
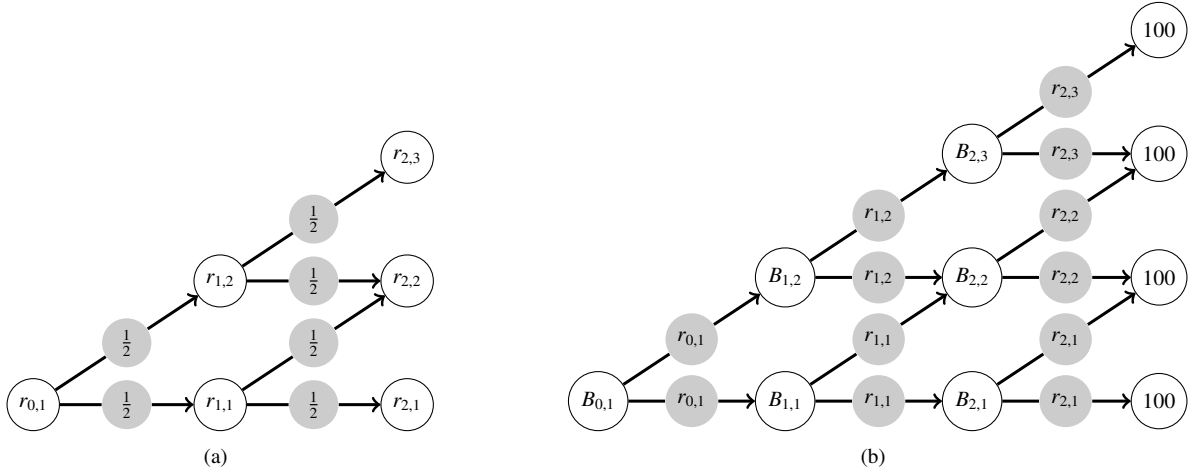

With the aim of simplifying the presentation, we consider that one period is equivalent to one year. Moreover, in whole paper we focus on the zero-coupon  bonds (zc-bonds). The corresponding modification
to shorter periods or use for bonds with coupons is straightforward. Figure \ref{figure:bdtp} presents the BDT model applied with maturity in $ T = 3 $ years. In Figure \ref{figure:bdtp} (a) the interest rate tree is considered where its values $r_ {i,j} $ for $ i =0,1,2$ and $ j = 1,\ldots,i+1$ need to be calibrated by the use of yields and volatilities data. Through their estimation, it is possible to calibrate the price of the bond by using the interest rate discount technique.
In Figure \ref{figure:bdtp} (b), we present the tree corresponding to the prices of a zc-bond for the Figure  \ref{figure:bdtp} (a). We denote by $B_{i,j}$ the zc-bond price corresponding to the period $i$ and state $j$.
For simplicity, we assume $B_{T,j}=100$, for $j=1,\dots,T+1$ that corresponds to the bond's face value (FV).

\subsection{The Zero Black-Derman-Toy model}\label{subsection:zbdt}

Our modification of the classical BDT interest rate tree model 
consists of adding to the dynamics the possibility of a downwards jump to a practically zero interest rate with a small probability at each time step. 
More precisely, in the ZBDT model the nodes of the form $(i,1)$ add a third possible downwards jump with a small probability $p$.
If this downwards jump is realized, then the process enters the so called Zero Interest Rate Policy Zone (ZIRP Zone), 
meaning that  interest rate becomes a small value $x_0$ 
(close to the target of the monetary policy). The other two transition probabilities (up and down)
for nodes of the form $(i,1)$ are equal to $\hat{p}=(1-p)/2$. When the process is in the ZIRP zone, 
 it remains there at each time step with a high probability $(1-q)$ and exits with probability $q$. For more information see \cite{zbdt}. It is important to note that $p$, $q$, $x_{0}$ have a clear interpretation as a probability of the crisis (in the basic period of time, which in this paper is $1$ year), a conditional probability of economic recovery from the financial crisis (in the basic period of time) and an assumed value of the interest rate in the ZIRP zone respectively. Moreover the ZBDT calibration uses the same data and the same hypothesis of independence between the periods as in the case of the BDT model. Similarly as in the BDT model it is also assumed that volatility depends only on the time.

\begin{figure}[H]
\centering
\subfigure[]{
\begin{tikzpicture}[scale=0.8, transform shape]
\tikzstyle{every node} = [circle, fill=gray!0,draw]

\node (22) at (3,2) {$r_{1,2}$};
\node (32) at (6,2) {$r_{2,2}$};
\node (33) at (6,4) {$r_{2,3}$};

\tikzstyle{every node} = [circle, fill=red!40,draw]
\node (11) at (0,0) {$r_{0,1}$};
\node (21) at (3,0) {$r_{1,1}$};
\node (31) at (6,0) {$r_{2,1}$};
\tikzstyle{every node} = [circle, fill=gray!40]

\node (2232) at (4.5,2) {$\frac12$};
\node (2233) at (4.5,3) {$\frac12$};
\foreach \from/\to in {11/1121,11/1122}
\draw [-,line width=0.4mm,fill=black] (\from) -- (\to);
\foreach \from/\to in {1121/21,1122/22}
\draw [->,line width=0.4mm,fill=black] (\from) -- (\to);
\foreach \from/\to in {21/2131,21/2132,22/2232,22/2233}
\draw [-,line width=0.4mm,fill=black] (\from) -- (\to);
\foreach \from/\to in {2131/31,2132/32,2232/32,2233/33}
\draw [->,line width=0.4mm,fill=black] (\from) -- (\to);
\tikzstyle{every node} = [circle, fill=blue!20,draw]
\node (20) at (3,-3) {$x_0$};
\node (30) at (6,-3) {$x_0$};
\tikzstyle{every node} = [circle, fill=blue!50]

\node (2030) at (4.5,-3) {$\scriptstyle1-q$};

\node (2031) at (4,-2) {$q$};
\tikzstyle{every node} = [circle, fill=red!60,draw]
\node (1120) at (1.5,-1.5) {$p$};
\node (2130) at (4,-1) {$p$};
\node (1121) at (1.5,0) {$\hat{p}$};
\node (1122) at (1.5,1) {$\hat{p}$};
\node (2131) at (4.5,0) {$\hat{p}$};
\node (2132) at (4.5,1) {$\hat{p}$};
\foreach \from/\to in {11/1120}
\draw [-,line width=0.4mm,fill=black] (\from) -- (\to);
\foreach \from/\to in {1120/20}
\draw [->,line width=0.4mm,fill=black] (\from) -- (\to);
\foreach \from/\to in {20/2030}
\draw [-,line width=0.4mm,fill=black] (\from) -- (\to);
\foreach \from/\to in {2030/30}
\draw [->,line width=0.4mm,fill=black] (\from) -- (\to);
\foreach \from/\to in {20/2031,21/2130}
\draw [-,line width=0.4mm,fill=black] (\from) -- (\to);
\foreach \from/\to in {2130/30,2031/31}
\draw [->,line width=0.4mm,fill=black] (\from) -- (\to);
\end{tikzpicture}
}\hspace{15mm}
\centering
\subfigure[]{
\begin{tikzpicture}[scale=0.8, transform shape]
\tikzstyle{every node} = [circle, fill=gray!0,draw]

\node (22) at (3,2) {$B_{1,2}$};

\node (32) at (6,2) {$B_{2,2}$};
\node (33) at (6,4) {$B_{2,3}$};

\node (42) at (9,2) {$100$};
\node (43) at (9,4) {$100$};
\node (44) at (9,6) {$100$}; 

\tikzstyle{every node} = [circle, fill=red!40,draw]
\node (21) at (3,0) {$B_{1,1}$};
\node (31) at (6,0) {$B_{2,1}$};
\node (41) at (9,0) {$100$};
\node (11) at (0,0) {$B_{0,1}$};
\tikzstyle{every node} = [circle, fill=red!60]
\node (1121) at (1.5,0) {$r_{0,1}$};
\node (1122) at (1.5,1) {$r_{0,1}$};
\node (2131) at (4.5,0) {$r_{1,1}$};
\node (2132) at (4.5,1) {$r_{1,1}$};

\node (3141) at (7.5,0) {$r_{2,1}$};
\node (3142) at (7.5,1) {$r_{2,1}$};
\node (3140) at (7,-1) {$r_{2,1}$};
\node (2130) at (4,-1) {$r_{1,1}$};
\node (1120) at (1.5,-1.5) {$r_{0,1}$};

\tikzstyle{every node} = [circle, fill=gray!40]
\node (3242) at (7.5,2) {$r_{2,2}$};
\node (3243) at (7.5,3) {$r_{2,2}$};
\node (3343) at (7.5,4) {$r_{2,3}$};
\node (3344) at (7.5,5) {$r_{2,3}$};
\node (2232) at (4.5,2) {$r_{1,2}$};
\node (2233) at (4.5,3) {$r_{1,2}$};
\foreach \from/\to in {11/1121,11/1122}
\draw [-,line width=0.4mm,fill=black] (\from) -- (\to);
\foreach \from/\to in {1121/21,1122/22}
\draw [->,line width=0.4mm,fill=black] (\from) -- (\to);
\foreach \from/\to in {21/2131,21/2132,22/2232,22/2233}
\draw [-,line width=0.4mm,fill=black] (\from) -- (\to);
\foreach \from/\to in {2131/31,2132/32,2232/32,2233/33}
\draw [->,line width=0.4mm,fill=black] (\from) -- (\to);
\foreach \from/\to in {31/3141,31/3142,32/3242,32/3243,33/3343,33/3344}
\draw [-,line width=0.4mm,fill=black] (\from) -- (\to);
\foreach \from/\to in {3141/41,3142/42,3242/42,3243/43,3343/43,3344/44}
\draw [->,line width=0.4mm,fill=black] (\from) -- (\to);
\tikzstyle{every node} = [circle, fill=blue!20,draw]
\node (20) at (3,-3) {$B_{1,0}$};
\node (30) at (6,-3) {$B_{2,0}$};
\node (40) at (9,-3) {$100$};
\tikzstyle{every node} = [circle, fill=blue!50]
\node (2030) at (4.5,-3) {$ x_{0}$};
\node (3040) at (7.5,-3) {$ x_{0}$};

\node (2031) at (4,-2) {$x_{0}$};

\node (3041) at (7,-2) {$x_{0}$};
\foreach \from/\to in {11/1120}
\draw [-,line width=0.4mm,fill=black] (\from) -- (\to);
\foreach \from/\to in {1120/20}
\draw [->,line width=0.4mm,fill=black] (\from) -- (\to);
\foreach \from/\to in {20/2030}
\draw [-,line width=0.4mm,fill=black] (\from) -- (\to);
\foreach \from/\to in {2030/30}
\draw [->,line width=0.4mm,fill=black] (\from) -- (\to);
\foreach \from/\to in {30/3040}
\draw [-,line width=0.4mm,fill=black] (\from) -- (\to);
\foreach \from/\to in {3040/40}
\draw [->,line width=0.4mm,fill=black] (\from) -- (\to);
\foreach \from/\to in {20/2031,21/2130,30/3041,31/3140}
\draw [-,line width=0.4mm,fill=black] (\from) -- (\to);
\foreach \from/\to in {2130/30,2031/31,3140/40,3041/41}
\draw [->,line width=0.4mm,fill=black] (\from) -- (\to);
\end{tikzpicture}
}
\caption{The ZBDT interest rate tree (a) and the corresponding tree of zc-bond prices (b) with $T=3$ and $FV=100$.}\label{figure:zbdtp}
\end{figure}
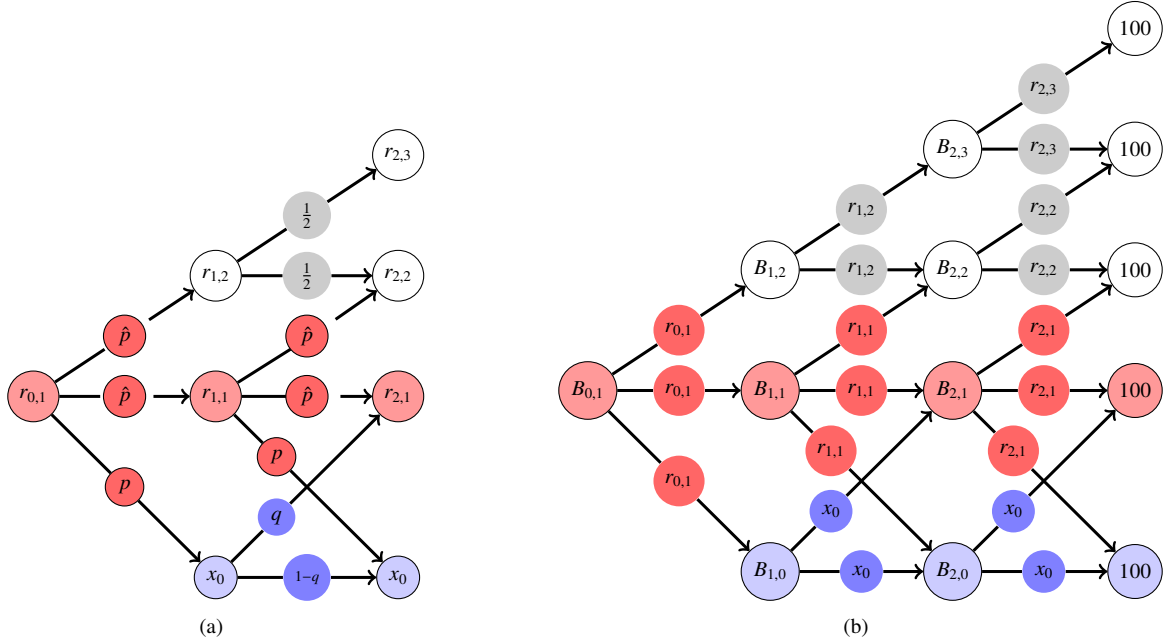

In Figure \ref{figure:zbdtp} the interest rate tree (a) and zc-bond price tree (b) for the ZBDT model are presented. Analogous to BDT model, we denote by $B_{i,j}$ the zc-bond price corresponding to the period $i$ and state $j$. For simplicity, we assume $B_{T,j}=100$, for $j=0,\dots,T+1$ that correspond to the bond's face value (FV). In this Figure we differentiate $3$ zones. In each zone, different dynamics is conserved what by different transition probabilities is manifested. The probabilities $p$ and $q$ have no influence on the first zone (white nodes with grey transition probabilities), therefore the dynamics of this zone remains the same as in the case of BDT model (i.e. the probability of transition is $u=d=1/2$). This  zone
comprises the range from the $3$rd  bottom row to the top row (labeled $(i,j)$ with $j=2,\ldots,T$ and $i=1,\ldots,T-1$). In the second zone (red nodes and the transition probabilities, labeled $(i,1)$ with $i=0,\ldots,T-1$) the transition probabilities corresponding to go up and down have the same value $\hat{p}=(1-p)/2$ and the model includes a small probability $p$ of falling into ZIRP zone. 
The last zone is the ZIRP-zone  (blue nodes and the transition probabilities, the nodes are marked as $x_{0}$). As mentioned before its transition probabilities are $q$ for a jump out of the ZIRP zone and $1-q$ for remaining there.

\subsection{The valuation of financial derivatives}\label{subsection:valuation}

In this subsection we consider some aspects of the valuation of derivatives introduced in Section $2$. As an extension of the BDT model, the ZBDT model is a way to valuate the different financial derivatives on bonds. Hence, the calibration on the interest rate tree and the corresponding (future) states of its underlying instrument (i.e. bond) is based.


Firstly, we focus on pricing the European-style options (Table \ref{table:1}) on bonds in both models. At the beginning we take the payoff function of the derivative $f$  on the nodes situated at the option's expiry time $S$ whose underlying asset is a bond with maturity in time $T$. Then using the probability transitions  on each branch we proceed the backward induction, similarly as it was done in \cite{crr}. At the end we obtain $V_{0,1}$ which the price of the derivative (at time $t=0$) is. Note that the function $f$ could be any payoff function from Table \ref{table:1}.

 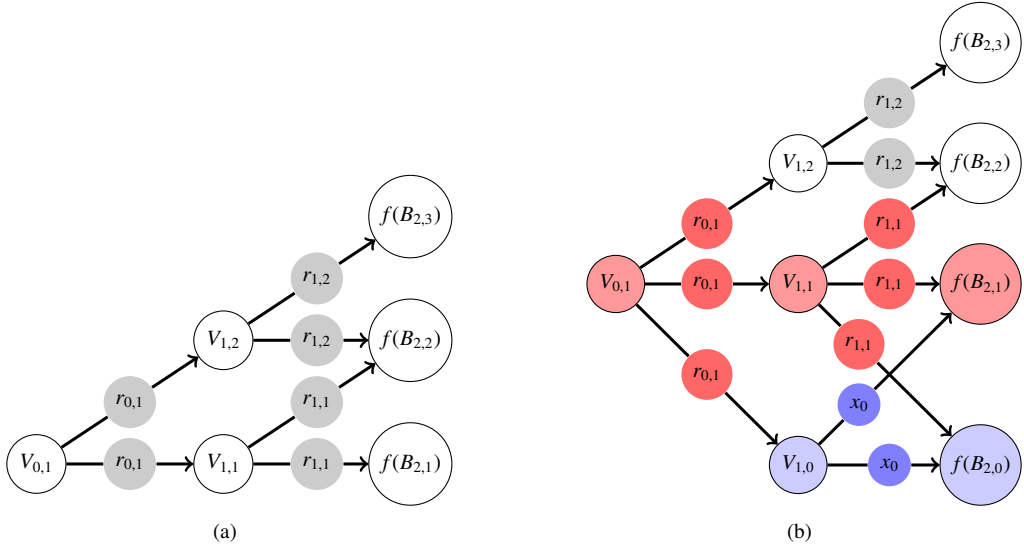
\begin{figure}
\centering 
\subfigure[]{
\begin{tikzpicture}[scale=0.82, transform shape]
\tikzstyle{every node} = [circle, fill=gray!0,draw]
\node (11) at (0,0) {$V_{0,1}$};
\node (21) at (3,0) {$V_{1,1}$};
\node (22) at (3,2) {$V_{1,2}$};
\node (31) at (6,0) {$f(B_{2,1})$};
\node (32) at (6,2) {$f(B_{2,2})$};
\node (33) at (6,4) {$f(B_{2,3})$};
\tikzstyle{every node} = [circle, fill=gray!40]
\node (1121) at (1.5,0) {$r_{0,1}$};
\node (1122) at (1.5,1) {$r_{0,1}$};
\node (2131) at (4.5,0) {$r_{1,1}$};
\node (2132) at (4.5,1) {$r_{1,1}$};
\node (2232) at (4.5,2) {$r_{1,2}$};
\node (2233) at (4.5,3) {$r_{1,2}$};
\foreach \from/\to in {11/1121,11/1122}
\draw [-,line width=0.4mm,fill=black] (\from) -- (\to);
\foreach \from/\to in {1121/21,1122/22}
\draw [->,line width=0.4mm,fill=black] (\from) -- (\to);
\foreach \from/\to in {21/2131,21/2132,22/2232,22/2233}
\draw [-,line width=0.4mm,fill=black] (\from) -- (\to);
\foreach \from/\to in {2131/31,2132/32,2232/32,2233/33}
\draw [->,line width=0.4mm,fill=black] (\from) -- (\to);
\end{tikzpicture}
}\hspace{15mm} 
\subfigure[]{
\begin{tikzpicture}[scale=0.8, transform shape]
\tikzstyle{every node} = [circle, fill=gray!0,draw]

\node (22) at (3,2) {$V_{1,2}$};

\node (32) at (6,2) {$f(B_{2,2})$};
\node (33) at (6,4) {$f(B_{2,3})$};

\tikzstyle{every node} = [circle, fill=red!40,draw]
\node (11) at (0,0) {$V_{0,1}$};
\node (21) at (3,0) {$V_{1,1}$};
\node (31) at (6,0) {$f(B_{2,1})$};
\tikzstyle{every node} = [circle, fill=gray!40]

\node (2232) at (4.5,2) {$r_{1,2}$};
\node (2233) at (4.5,3) {$r_{1,2}$};
\tikzstyle{every node} = [circle, fill=red!60]
\node (1121) at (1.5,0) {$r_{0,1}$};
\node (1122) at (1.5,1) {$r_{0,1}$};
\node (2131) at (4.5,0) {$r_{1,1}$};
\node (2132) at (4.5,1) {$r_{1,1}$};
\node (1120) at (1.5,-1.5) {$r_{0,1}$};
\node (2130) at (4,-1) {$r_{1,1}$};
\foreach \from/\to in {11/1121,11/1122}
\draw [-,line width=0.4mm,fill=black] (\from) -- (\to);
\foreach \from/\to in {1121/21,1122/22}
\draw [->,line width=0.4mm,fill=black] (\from) -- (\to);
\foreach \from/\to in {21/2131,21/2132,22/2232,22/2233}
\draw [-,line width=0.4mm,fill=black] (\from) -- (\to);
\foreach \from/\to in {2131/31,2132/32,2232/32,2233/33}
\draw [->,line width=0.4mm,fill=black] (\from) -- (\to);

\tikzstyle{every node} = [circle, fill=blue!20,draw]
\node (20) at (3,-3) {$V_{1,0}$};
\node (30) at (6,-3) {$f(B_{2,0})$};
\tikzstyle{every node} = [circle, fill=blue!50]

\node (2030) at (4.5,-3) {$x_{0}$};


\node (2031) at (4,-2) {$x_{0}$};

\foreach \from/\to in {11/1120}
\draw [-,line width=0.4mm,fill=black] (\from) -- (\to);
\foreach \from/\to in {1120/20}
\draw [->,line width=0.4mm,fill=black] (\from) -- (\to);
\foreach \from/\to in {20/2030}
\draw [-,line width=0.4mm,fill=black] (\from) -- (\to);
\foreach \from/\to in {2030/30}
\draw [->,line width=0.4mm,fill=black] (\from) -- (\to);

\foreach \from/\to in {20/2031,21/2130}
\draw [-,line width=0.4mm,fill=black] (\from) -- (\to);
\foreach \from/\to in {2130/30,2031/31}
\draw [->,line width=0.4mm,fill=black] (\from) -- (\to);
\end{tikzpicture}
}
\caption{Option's valuation for the  BDT model (a) and ZBDT model (b) with  $S=2$.
}\label{figure:bdtoption2}
\end{figure}


Finally let us consider the valuation of  the American (Vanilla) options
on bonds in both models.
At the beginning we proceed as in the case of European options, we take the payoff function on the nodes at time $S$ to obtain the values $V_{S,j}=f(B_{S,j})$, where $f$ is the payoff function (see Table $2$), and $j=1,\ldots,i+1$ or $j=0,\ldots,i+1$ for the BDT and ZBDT respectively. Using the probability transitions and the corresponding interest rates we obtain the values $V^{*}_{i,j}$, where $i=S-1$, $j=1,\ldots,i+1$ or $j=0,\ldots,i+1$ for the BDT and ZBDT respectively. 
 Then we have to verify if the immediate exercise of the option is not the most profitable strategy. So, for $i=S-1$, $j=1,\ldots,i+1$ (for BDT) or $j=0,\ldots,i+1$ (for ZBDT) we take $V_{i,j}=\max(V^{*}_{i,j},f_{V}(B_{i,j}))$, where $f_{V}$ is the payoff function of the corresponding American (Vanilla) option. We proceed in an iterative manner for $i=S-2,\ldots,0$. At the end, we get $V_{0,1}$ which is the price of the derivative (at time $t=0$).
 
 Note that in both option styles (European and American) we are valuing Vanilla and Barrier options. However in the case of the American Barrier options valuation, additionally we have to check if the node of the underlying instrument did not invalidate the contract by exceeding or not exceeding predetermined barrier(s). The European and American-style options valuation for both models is summarized in Figure \ref{figure:bdtoption2}.
The algorithms of  options valuation are presented in Algorithm $1$ and  Algorithm $2$ . In them, $f$ represents  the payoff function of the option and $f_{V}$ represents the payoff function of the corresponding American (Vanilla) option. Moreover: 
\begin{equation*}
    \forall\  j\ \text{\  i.e.}
    \begin{cases}
      j=1,\ldots,i+1 \text{\ for the BDT or\ } j=0,\ldots,i+1 \text{ for the ZBDT}, & \text{if}\ i>0 \\
      j=1 \text{ for both models}, & \text{if } i=0.
    \end{cases}
  \end{equation*}
It is important to note, that the price $B_{i,j}$ can invalidate the contract by exceeding barrier or interval of barriers in case of the knock-out options or by not exceeding barrier or interval of barriers in case of the knock-in options.

\begin{minipage}{0.46\textwidth}
\vspace{-2.2cm}
\begin{algorithm}[H]
    \caption{European-style option valuation}
    \begin{algorithmic}[1]
     \State $\forall j$ $V_{S,j}=f(B_{S,j})$
\For{$i$ from $S-1$ to $0$} $\forall j$
\State \text{By backward induction compute $V_{i,j}$,}
\EndFor\label{euclidendwhile}
\State \textbf{return} $V_{0,1}$.
    \end{algorithmic}
\end{algorithm}
\end{minipage}
\hfill
\begin{minipage}{0.46\textwidth}
\begin{algorithm}[H]
    \caption{American-style option valuation}\label{algorithm1}
    \begin{algorithmic}[1]
        \State $\forall j$ $V_{S,j}=f(B_{S,j})$
\For{$i$ from $S-1$ to $0$} $\forall j$
\State \text{By backward induction compute $V^{*}_{i,j}$,}
\If{$B_{i,j}$\text{ did not invalidate the contract} }
\State	$V_{i,j}=\max\left(f_{V}\left(B_{i,j}\right),V^{*}_{i,j}\right)$
\Else
\State $V_{i,j}=0$
\EndIf
\EndFor\label{euclidendwhile}
\State \textbf{return} $V_{0,1}$.
    \end{algorithmic}
\end{algorithm}
\end{minipage}\newline
\newline

\section{Comparison of BDT model and ZBDT model}\label{section:comparación}

In this section, the classical BDT model and the novel ZBDT model are calibrated in order to compare the performance in the valuation  of the Vanilla and Barrier options. In the work both models are compared through the options prices and the implied volatility.

In order to  compare both models, an "S-shaped" term structure (i.e. curve with parts decreasing and increasing) is used. The idea of this shape of the curve is that it has parts of the  three main types of yield curve shapes: normal (upward sloping curve), inverted (downward sloping curve) and flat. For the volatility in each yield value we consider a decreasing function because  usually it is observed in the sovereign debt. We analyse the case for the long term, so for this reason we select  one-year  nodes (frequency annually)  of the 10-year term structure ($T=10$). The values of yield ($\%$) and volatility ($\%$) can be found in equation (\ref{yield-vol}) below. For the ZBDT model, we assume  the parameters $x_{0}=0.25\%$, $p=0.02$, $q=0.01$. The input data and the set of  parameters we will call the "Example". In  Table \ref{tabla1_r_BDT} and Table \ref{tabla1_r_ZBDT}, we show the trees of the interest rate for BDT and ZBDT respectively. Table \ref{tabla1_B_BDT} and Table \ref{tabla1_B_ZBDT} contain the corresponding trees for the zc-bond prices. 

\begin{align}
\label{yield-vol}
yield &= (2.60,\  2.50,\ 2.47,\ 2.48,\ 2.49,\ 2.53,\ 2.61,\ 2.74,\ 2.92,\ 3.17); \nonumber\\
Vol&=(40.0,\ 34.0,\ 29.5,\ 28.9,\ 27.2,\ 26.0,\ 25.1,\ 24.2,\ 23.2,\ 23.1).
\end{align}

\begin{table}[h]
\centering
\scalebox{1}{
\begin{tabular}{|cccccccccc|}
\hline
      &         &         &          &          &          &          &          &          & 41.20\\
      &         &         &          &          &          &          &          & 19.00 & 25.32\\
      &         &         &          &          &          &          & 14.22 & 12.83 & 15.56\\
      &         &         &          &          &          & 10.30 & 9.34 & 8.66 & 9.56\\
      &         &         &          &          & 7.44 & 6.64 & 6.13 & 5.85 & 5.88\\
      &         &         &          &5.72 & 4.79 & 4.28 & 4.03 & 3.95 & 3.61\\
      &         &         & 5.34 & 3.64 & 3.08 & 2.76 & 2.65 & 2.67 & 2.22\\
      &         & 3.71 & 3.00 & 2.32 & 1.98 & 1.78 & 1.74 & 1.80 & 1.36\\
      & 3.19 & 2.28 & 1.68 & 1.48 & 1.28 & 1.15 & 1.14 & 1.22 & 0.84\\
2.60 & 1.62 & 1.41 & 0.94 & 0.94 & 0.82 & 0.74 & 0.75 & 0.82 & 0.52\\		
\hline
\end{tabular}}
\caption{BDT tree of interest rates ($\%$) for the Example.}\label{tabla1_r_BDT}
\end{table}

\begin{table}[h]
\centering
\scalebox{1}{
\begin{tabular}{|cccccccccc|}
\hline
      &         &         &         &         &         &         &         &         &77.00 \\
      &         &         &         &         &         &         &         & 33.79 & 40.79 \\
      &         &         &         &         &         &         & 26.30 & 19.56 & 21.60 \\
      &         &         &         &         &         & 18.66 & 14.15 & 11.32 & 11.44 \\
      &         &         &         &         & 12.65 & 9.51 & 7.61 & 6.55 & 6.06 \\
      &         &         &         & 8.94 & 6.29 & 4.84 & 4.09 & 3.79 & 3.21 \\
      &         &         & 7.47 & 4.32 & 3.13 & 2.47 & 2.20 & 2.20 & 1.70 \\
      &         & 4.77 & 3.10 & 2.08 & 1.55 & 1.26 & 1.18 & 1.27 & 0.90 \\
      & 3.60 & 2.08 & 1.29 & 1.00 & 0.77 & 0.64 & 0.64 & 0.74 & 0.48 \\
2.60 & 1.32 & 1.06 & 0.57 & 0.51 & 0.40 & 0.33 & 0.35 & 0.45 & 0.26 \\
      & 0.25 & 0.25 & 0.25 & 0.25 & 0.25 & 0.25 & 0.25 & 0.25 & 0.25 \\
\hline
\end{tabular}}
\caption{ZBDT tree of interest rates ($\%$) for the Example.}\label{tabla1_r_ZBDT}
\end{table}

\begin{table}[h]
\centering
\scalebox{1}{
\begin{tabular}{|ccccccccccc|}
\hline
          &          &          &          &          &          &          &          &          &           & 100 \\
          &          &          &          &          &          &          &          &          & 70.82  & 100 \\
          &          &          &          &          &          &          &          & 63.28 & 79.80  & 100 \\
          &          &          &          &          &          &          & 59.97 & 73.71 & 86.54  & 100 \\
          &          &          &          &          &          & 59.43 & 71.12 & 81.82 & 91.27  & 100 \\
          &          &          &          &          & 60.60 & 70.80 & 79.87 & 87.73 & 94.45  & 100 \\
          &          &          &          & 62.62 & 71.80 & 79.68 & 86.32 & 91.85 & 96.51  & 100 \\
          &          &          & 64.59 & 73.46 & 80.46 & 86.20 & 90.85 & 94.65 & 97.83  & 100 \\
          &          & 67.46 & 75.33 & 81.72 & 86.77 & 90.78 & 93.94 & 96.50 & 98.65  & 100 \\
          & 70.26 & 77.54 & 83.30 & 87.68 & 91.18 & 93.90 & 96.02 & 97.72 & 99.17  & 100 \\
 73.19 & 79.93 & 84.91 & 88.91 & 91.81 & 94.18 & 96.00 & 97.39 & 98.52 & 99.49  & 100 \\
\hline
\end{tabular}}
\caption{BDT tree of zc-bond prices for the  Example.}\label{tabla1_B_BDT}
\end{table}

\begin{table}[h]
\centering
\scalebox{1}{
\begin{tabular}{|ccccccccccc|}
\hline    &          &          &          &          &          &          &          &          &          & 100 \\
          &          &          &          &          &          &          &          &          & 56.50 & 100 \\
          &          &          &          &          &          &          &          & 47.66 & 71.03 & 100 \\
          &          &          &          &          &          &          & 44.24 & 64.10 & 82.24 & 100 \\
          &          &          &          &          &          & 44.73 & 61.91 & 77.24 & 89.73 & 100 \\
          &          &          &          &          & 47.80 & 62.97 & 76.01 & 86.35 & 94.29 & 100 \\
          &          &          &          & 52.19 & 65.91 & 77.13 & 85.72 & 92.10 & 96.89 & 100 \\
          &          &          & 56.68 & 69.64 & 79.39 & 86.61 & 91.78 & 95.51 & 98.33 & 100 \\
          &          & 62.16 & 73.57 & 82.06 & 88.14 & 92.41 & 95.37 & 97.48 & 99.11 & 100 \\
          & 67.45 & 77.60 & 84.86 & 89.86 & 93.37 & 95.78 & 97.42 & 98.59 & 99.53 & 100 \\
 73.19 & 81.82 & 87.58 & 91.77 & 94.48 & 96.40 & 97.69 & 98.56 & 99.20 & 99.75 & 100 \\
          & 97.53 & 97.87 & 98.18 & 98.47 & 98.74 & 99.00 & 99.25 & 99.50 & 99.75 & 100 \\
          \hline
\end{tabular}}
\caption{ZBDT tree of zc-bond prices for the  Example.}\label{tabla1_B_ZBDT}
\end{table}

\begin{figure}[h]

\begin{center}
\subfigure[]{
\includegraphics[width=0.31
\textwidth]{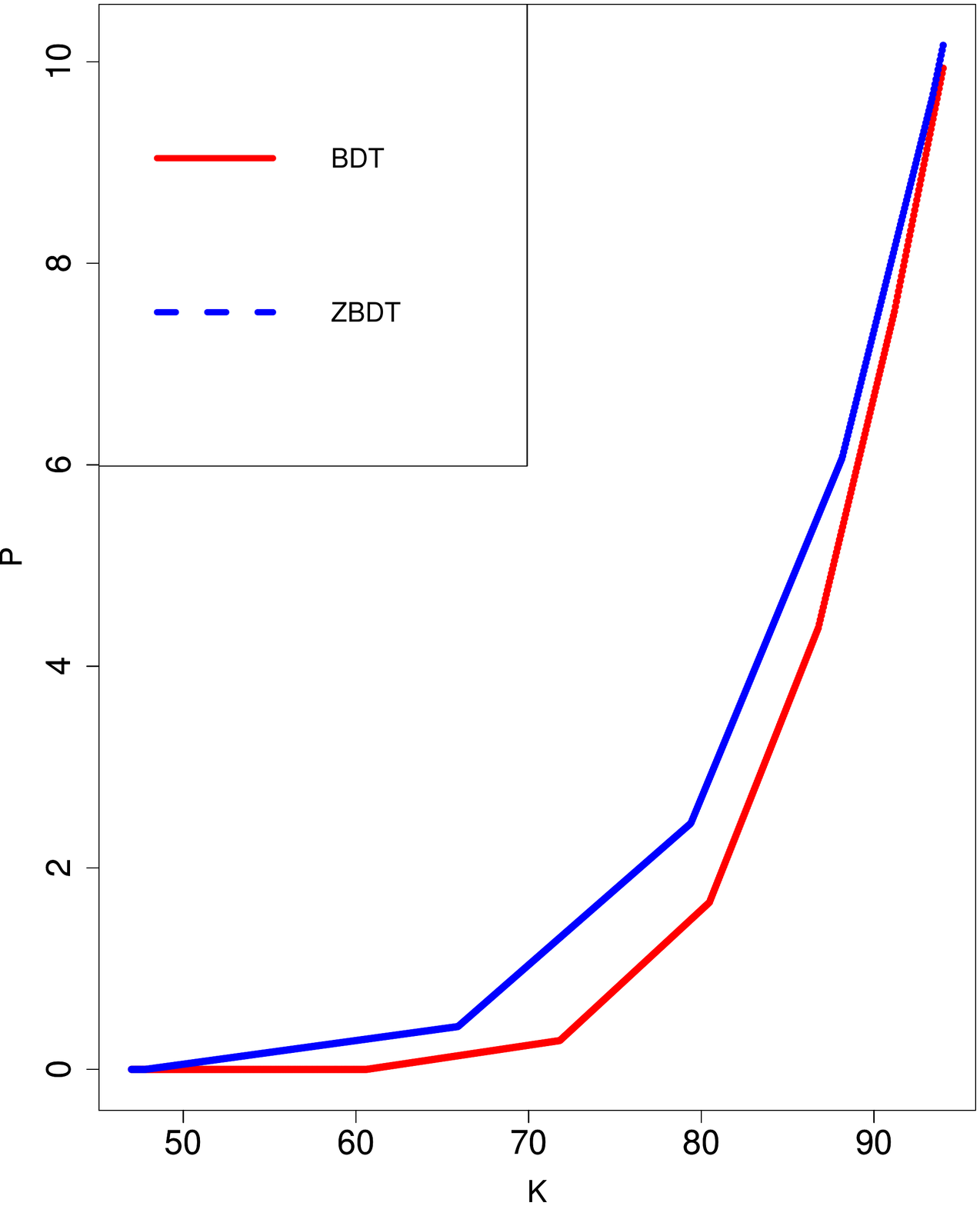}
}
\subfigure[]{
\includegraphics[width=0.31\textwidth]{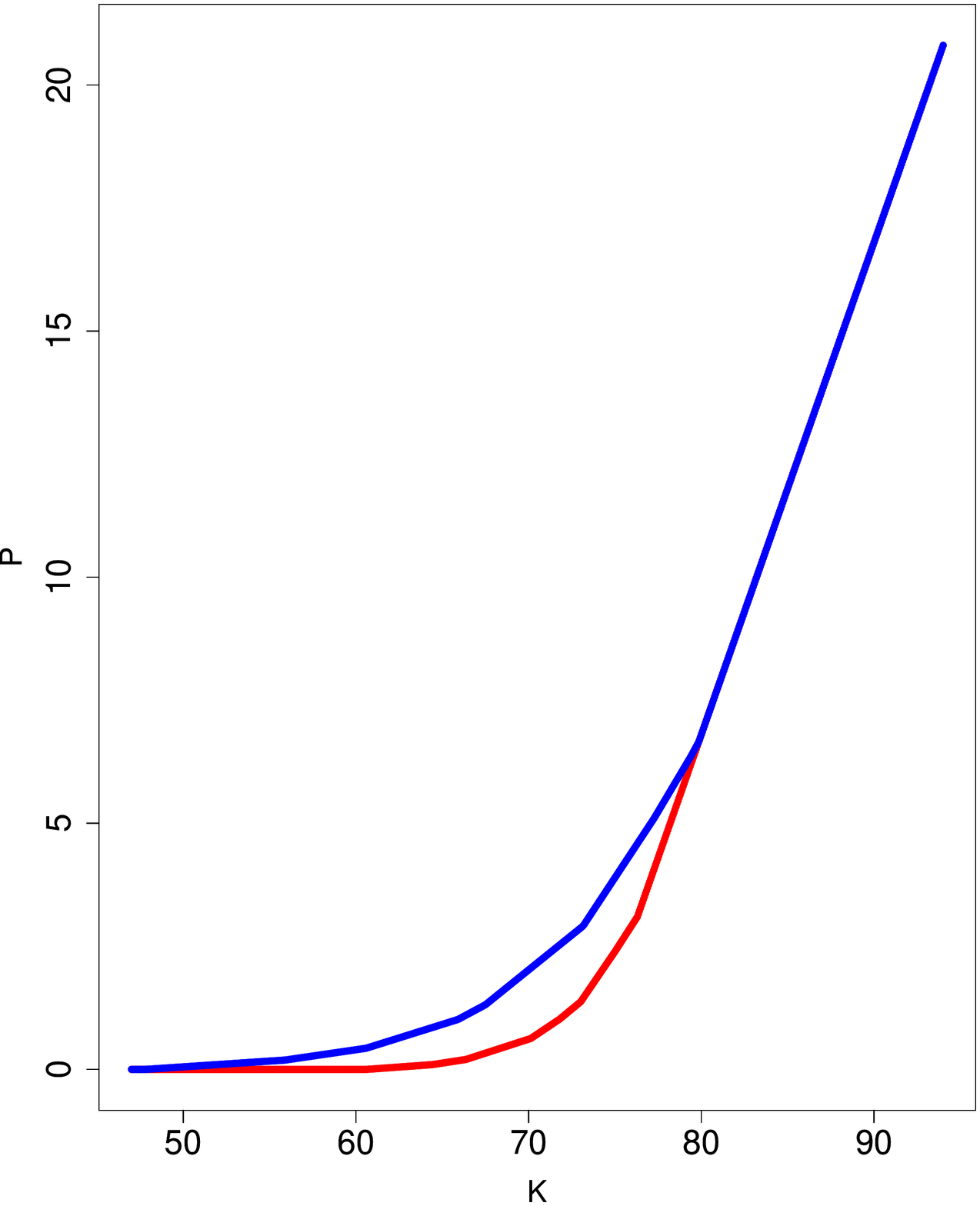}
}
\subfigure[]{
\includegraphics[width=0.31\textwidth]{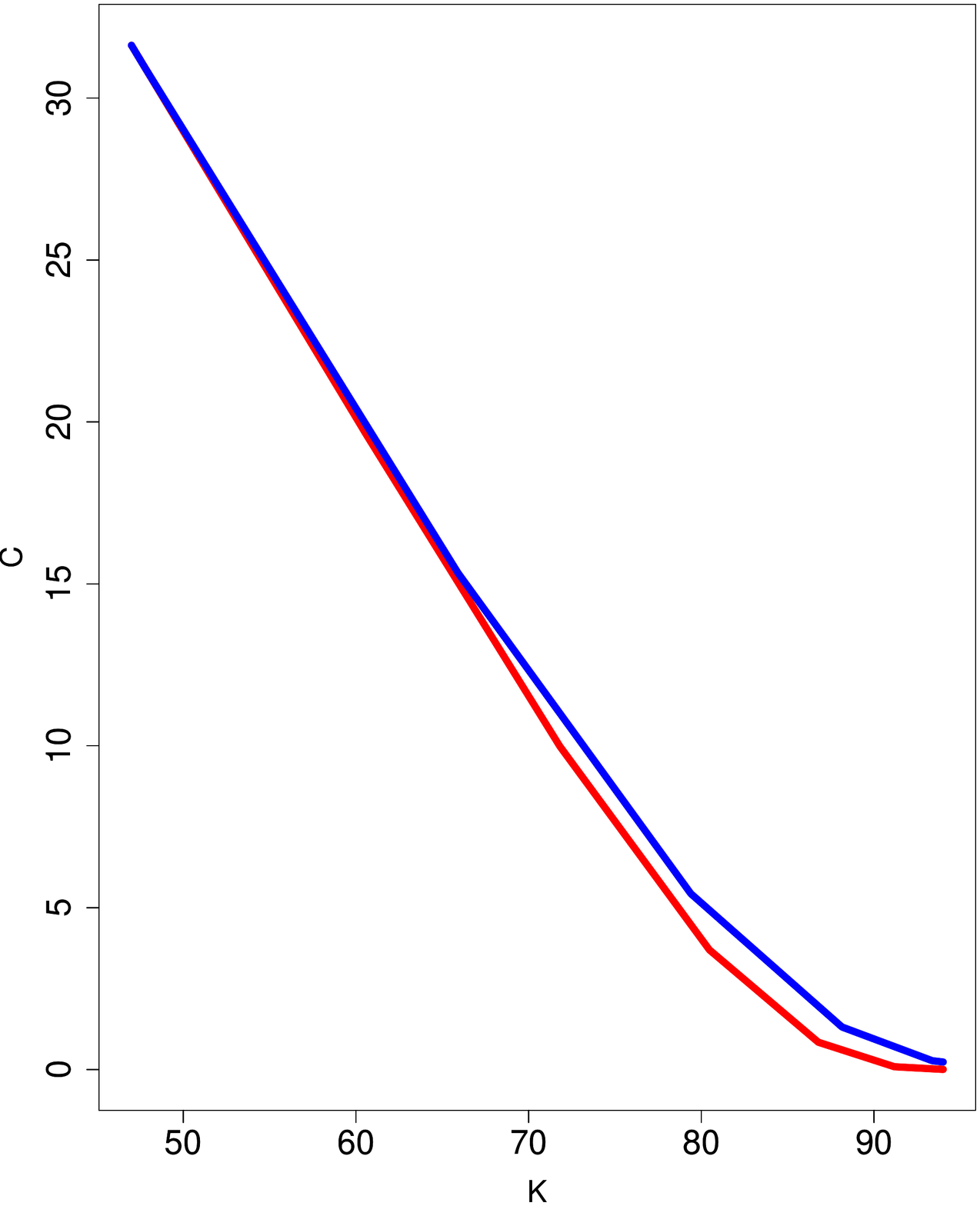}
}
\caption{Prices of Vanilla options on bonds from Table \ref{tabla1_B_BDT} and Table \ref{tabla1_B_ZBDT} in dependence of the strike price $K$ for ZBDT (blue) and BDT model (red).}\label{implvolVan}
\end{center}
\end{figure}

\begin{figure}[h]

\begin{center}
\subfigure[]{
\includegraphics[width=0.235\textwidth]{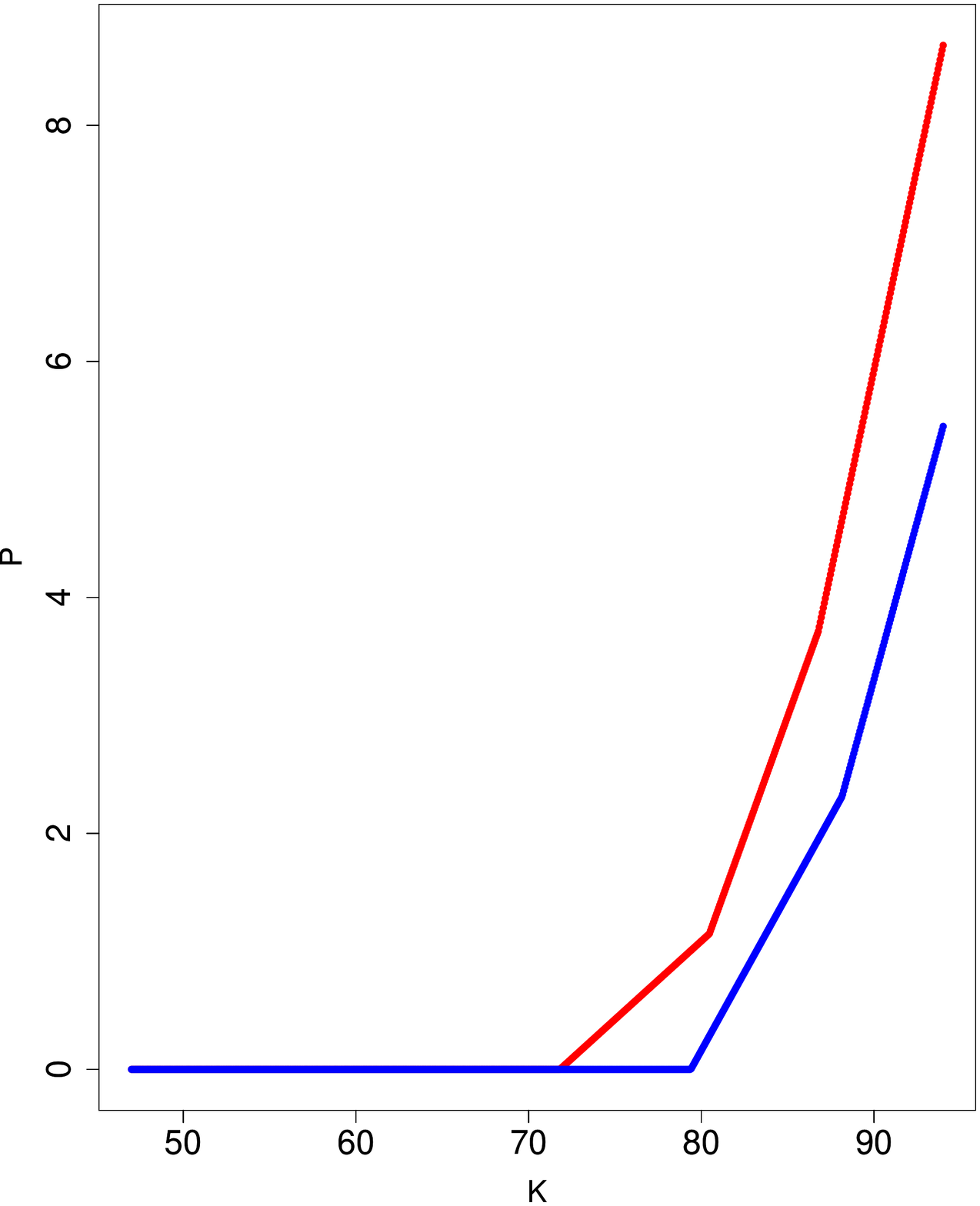}
}
\subfigure[]{
\includegraphics[width=0.235\textwidth]{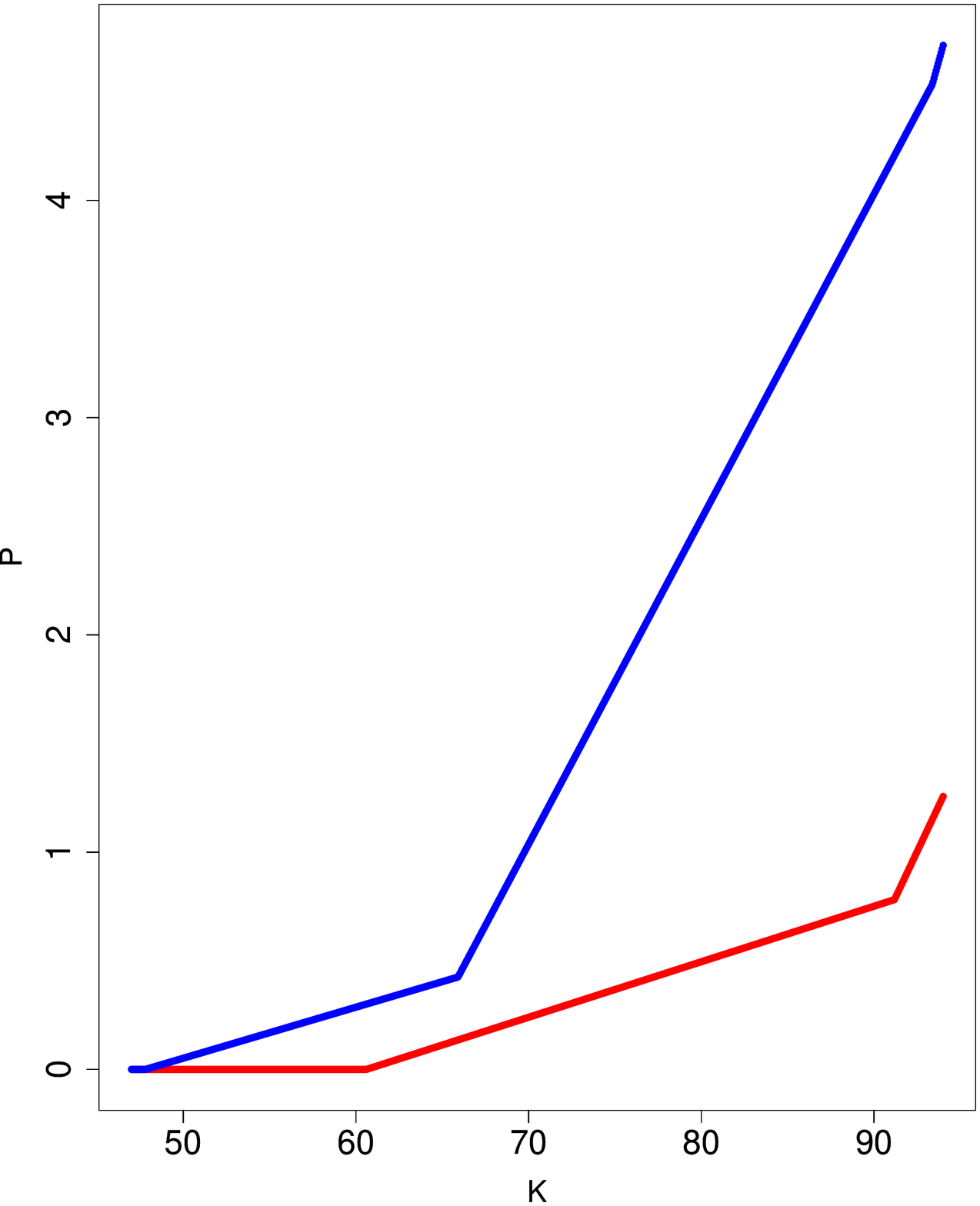}
}
\subfigure[]{
\includegraphics[width=0.235\textwidth]{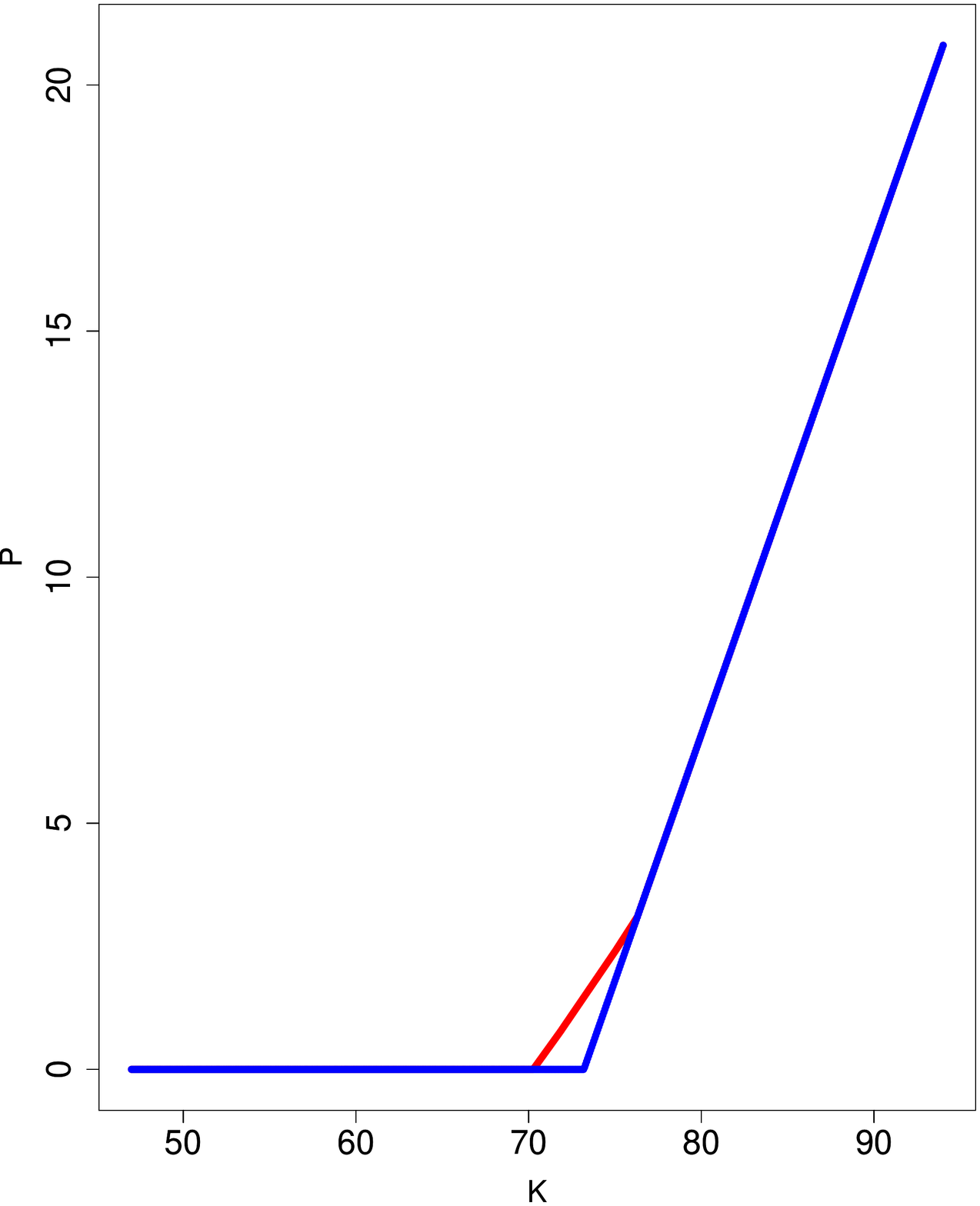}
}
\subfigure[]{
\includegraphics[width=0.235\textwidth]{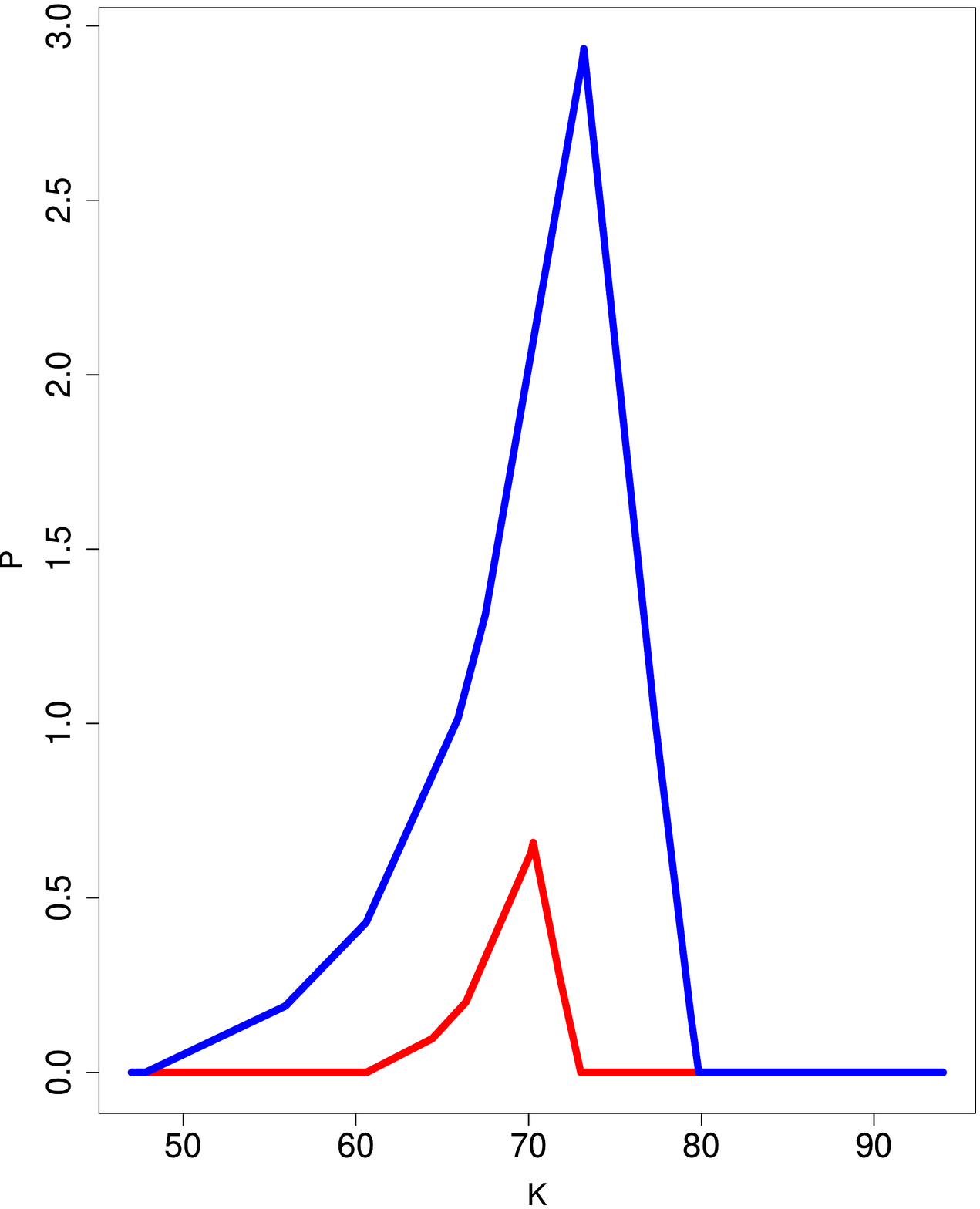}
}
\newline
\subfigure[]{
\includegraphics[width=0.235\textwidth]{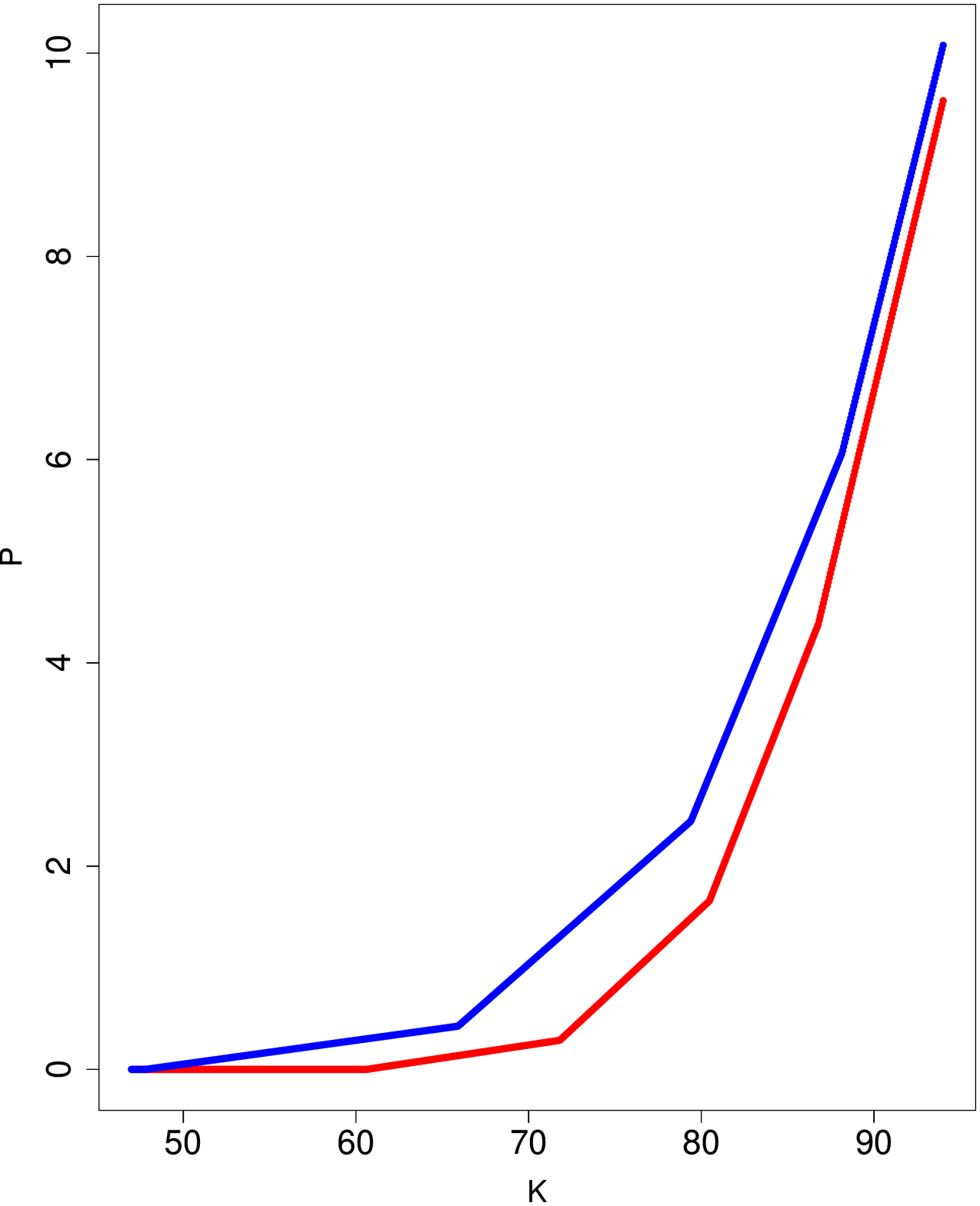}
}
\subfigure[]{
\includegraphics[width=0.235\textwidth]{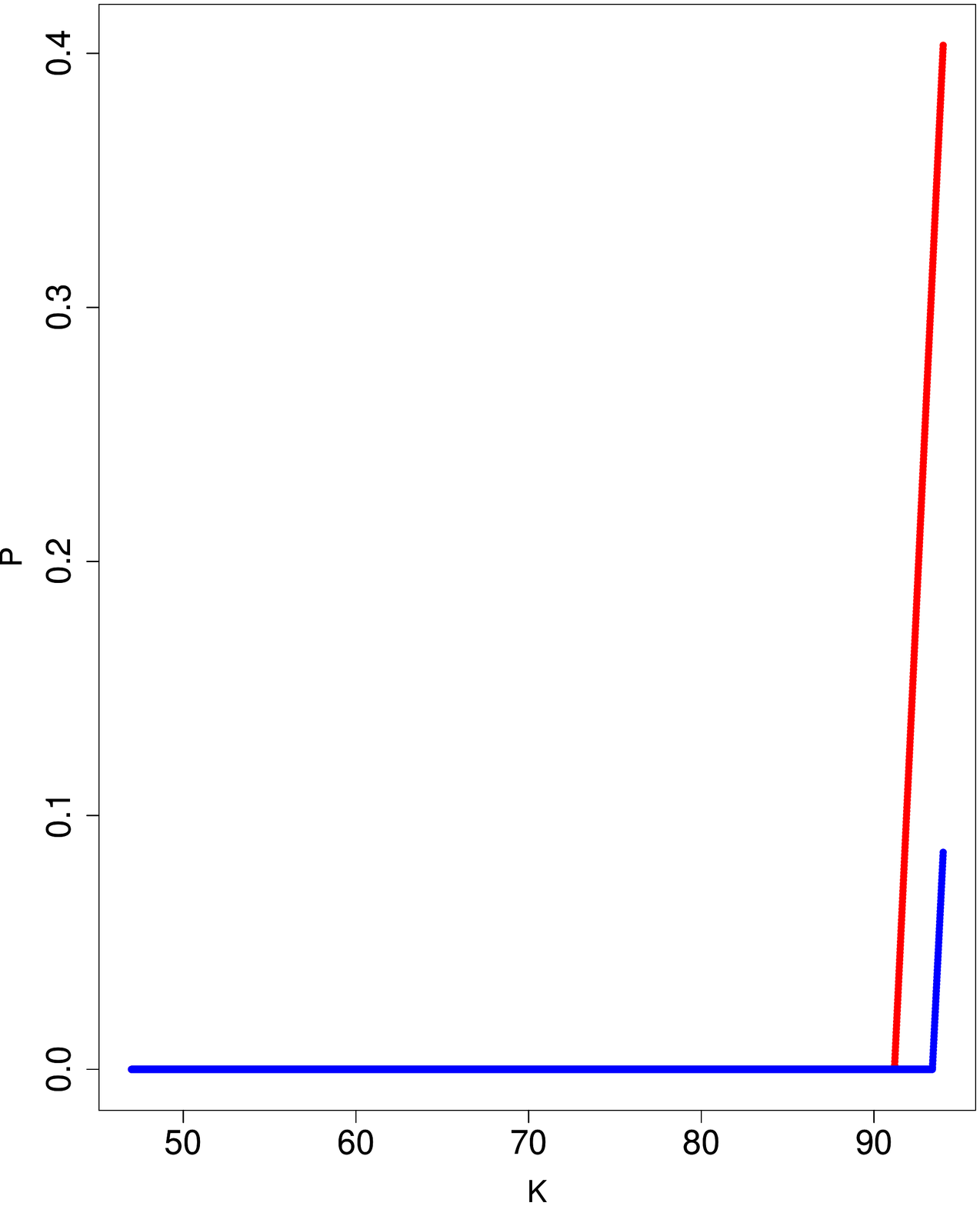}
}
\subfigure[]{
\includegraphics[width=0.235\textwidth]{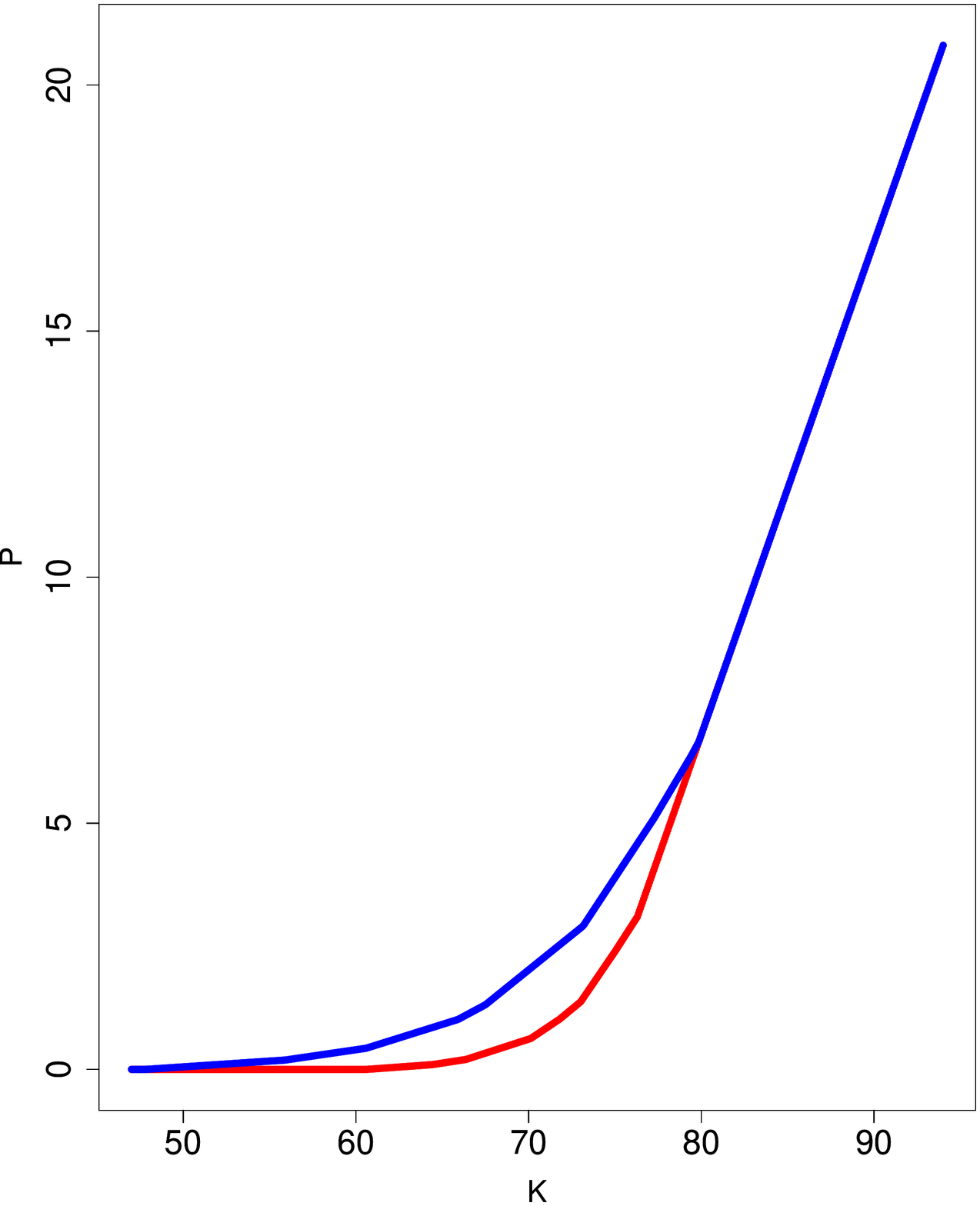}
}
\subfigure[]{
\includegraphics[width=0.235\textwidth]{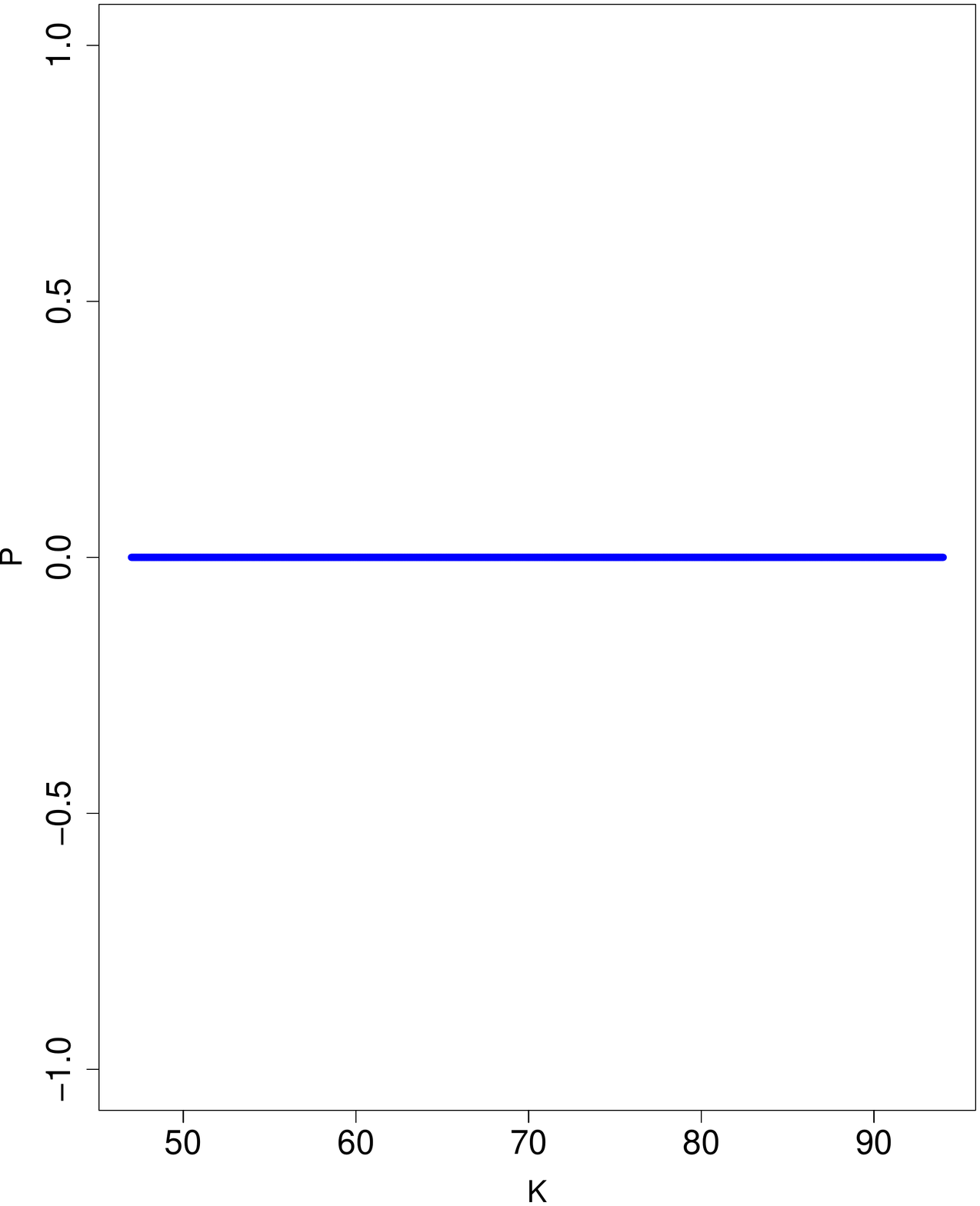}
}
\caption{Option prices in dependence of the strike price $K$ for ZBDT model (blue) and BDT model (red) for the Barrier options on bonds from Table \ref{tabla1_B_BDT} and Table \ref{tabla1_B_ZBDT}.}\label{implvolsec5}
\end{center}
\end{figure}

From the trees of the zc-bonds prices in both models (Table \ref{tabla1_B_BDT} and Table \ref{tabla1_B_ZBDT}) it is possible to valuate the financial derivatives introduced in Section \ref{seccion:2}. We use selected  derivatives written at $t=0$ with exercise time five years ($S=5$). In Figure \ref{implvolVan} and Figure \ref{implvolsec5}, we present the option prices in dependence of the strike price $K$. Here and in the rest of the work the red colour corresponds to the BDT model and the blue colour to the ZBDT model. In Figure \ref{implvolVan} we show the prices of
(a) European put, (b) American put,  c) European call option. The non-zero probability of the fall into ZIRP zone causes the prices of the zc-bonds increase. As the consequence the prices of the Vanilla options in ZBDT are higher than in the BDT model. It follows our aims, because the risk related to the Vanilla options put is higher for the ZBDT than for the BDT model.

\begin{figure}[h!]
\begin{center}
\subfigure[]{
\includegraphics[width=0.31\textwidth]{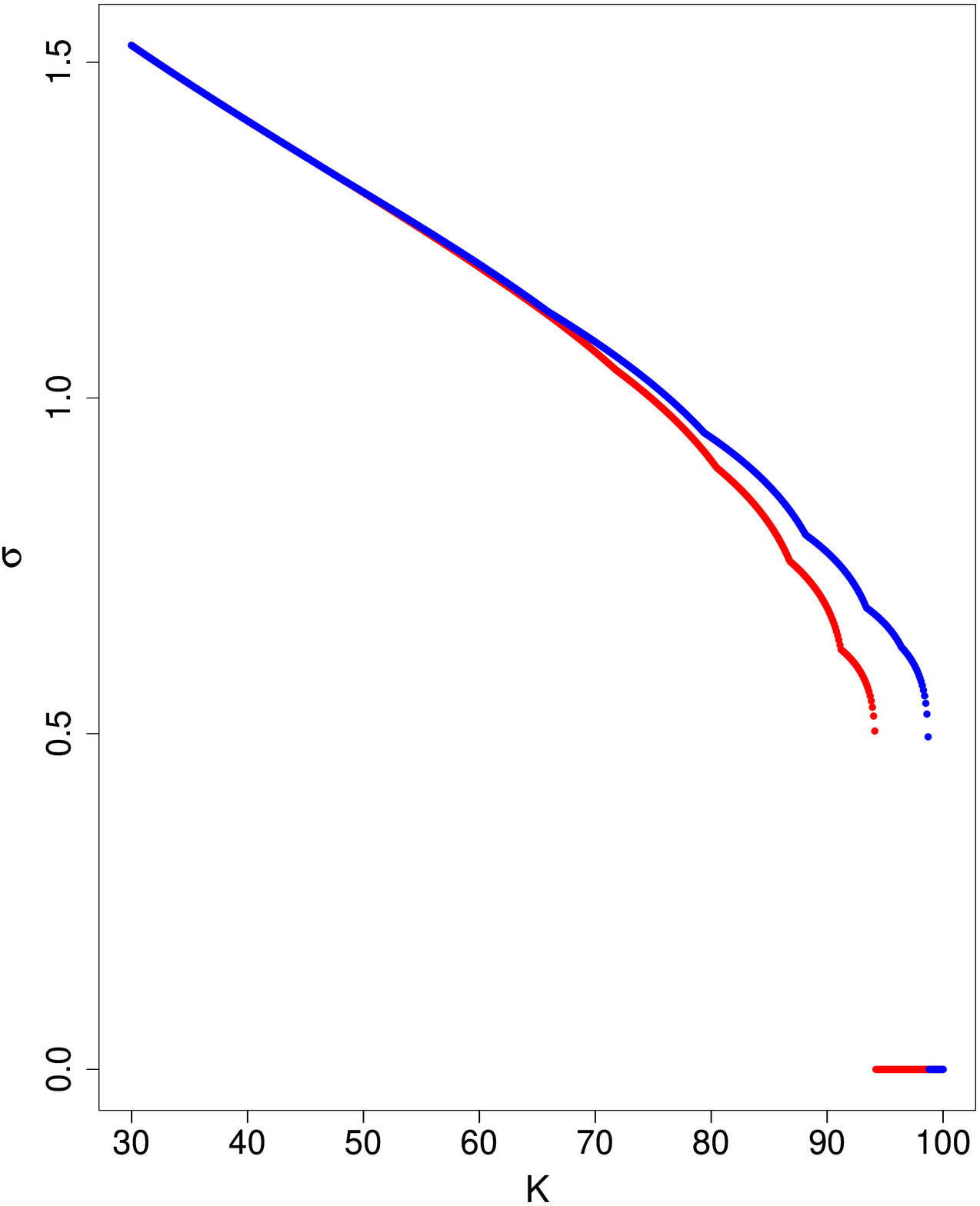}
}
\subfigure[]{
\includegraphics[width=0.31\textwidth]{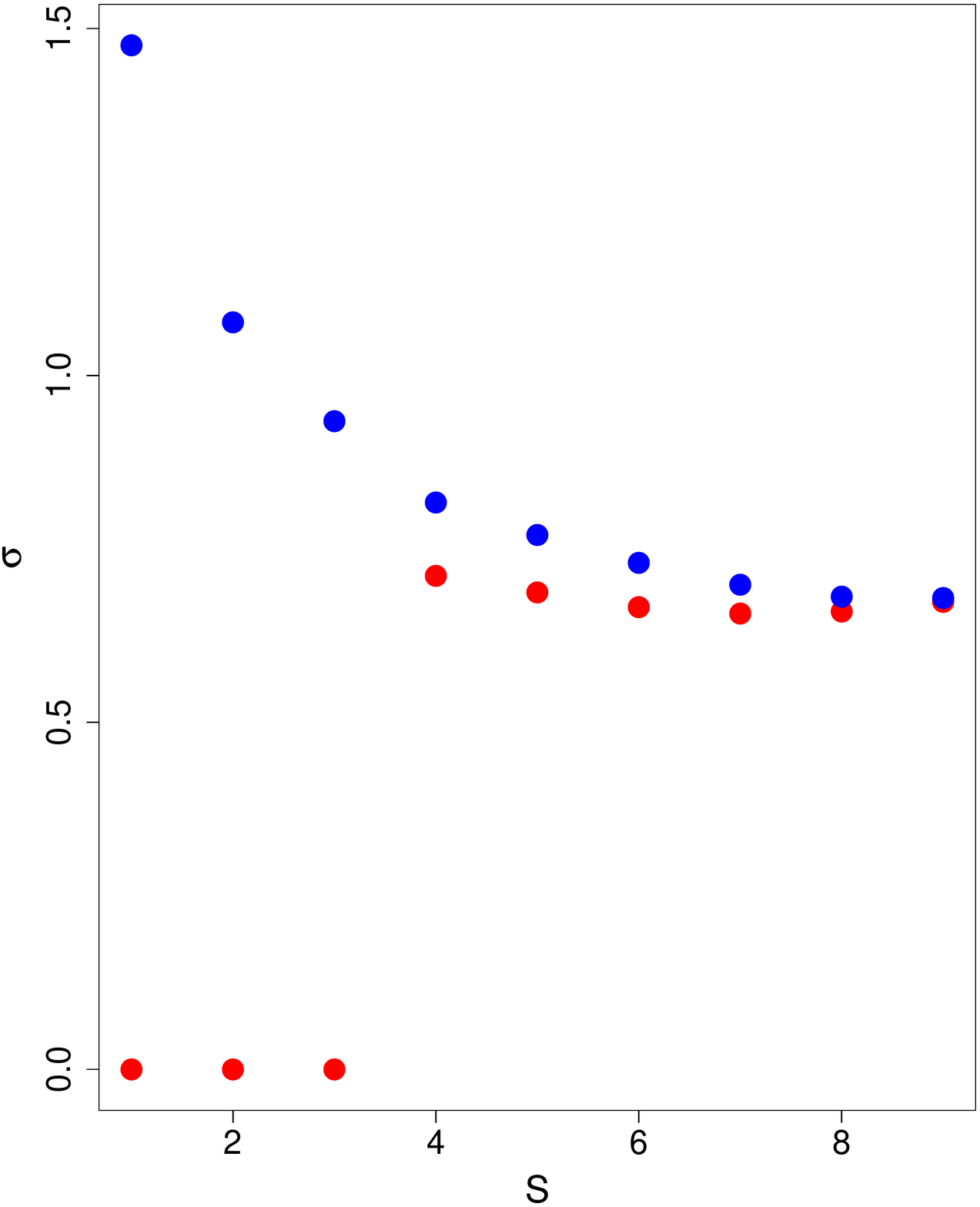}
}
\caption{Implied volatility for the Example: (a)  in dependence of the strike price $K$, (b) in dependence of  expiry time $S$ of the option for $K=90$.}\label{ImpVol}
\end{center}
\end{figure}

In the case of the European (Vanilla) options valuation and in the modelling risk related to these contracts it is important to analyse the implied volatility. The implied volatility at time $t$ of an European Vanilla option written on a zc-bond that matures at time $T$, 
with strike price $K$ and expiry time $S$ ($t<S<T$), is computed via Black's formula \cite{black},
which states that the price of the European call option is
\begin{equation*}
C=B(t,T)\Phi(d_1)-KB(t,S)\Phi(d_2),
\end{equation*}
where
$$
d_{1,2}={\log\left({B(t,T)\over KB(t,S)}\right)\over\sigma\sqrt{S-t}}\pm{\sigma\sqrt{S-t}\over 2};
$$
and  $\Phi$ is the cumulative normal distribution function. Analogous formula for the put option through put-call parity can be obtained. Note that regardless of the model under consideration, we assume that the implied volatility is equal to $0$ if the corresponding option is worthless. For more details concerning the estimation of the implied volatility, see  \cite{mcd}.   In Figure \ref{ImpVol}, we show the implied volatility for prices of the European Vanilla call option for the Example. The  panel (a) concerns the dependence between the implied volatility $\sigma$ and the strike price $K$. In both models implied volatility by similar curves is expressed. The  panel (b) concerns the dependence between the implied volatility $\sigma$ and the expiry time of the option $S$ for $K=90$. The implied volatility in the BDT model for $S=2$ and $S=3$ is assumed to be equal to $0$. When analyzing these results, it is concluded that the ZBDT model can be useful. The first  reason is including the existence of a possible jump up in the price of the zc-bond. This property follows that the price of those derivatives must be higher for ZBDT than for BDT model. The second reason is the existence of (feasible) cases that the classical BDT model valuates the derivative with a price equal to $0$, since the corresponding price in the ZBDT model is positive . Both results confirm that the risk manifested by the implied volatility is higher for the ZBDT model than for BDT model. 

The prices of Barrier options are collected in Figure \ref{implvolsec5}.
The left panels are calculated for European put options: (a) Knock-Double-Out,  (b) Knock-Double-In, (e) Knock-Up-Out, (f) Knock-Up-In. The right panels (c, d, g, h) represent the prices of American put options having the same orders as their European equivalents. The barriers considered in the Example are $H^{-}=70$, $H^{+}=90$.  We can observe that depending on the barrier contract, different relations between the option prices of BDT and ZBDT model can be obtained. The curve can have different shape depending on the model and type of the option. Therefore, the price corresponding to ZBDT does not have to be higher than in BDT model. Since the risk corresponding to the Barrier options is lower than for its Vanilla analogs (for issuer of the option), the same relation between the prices of the instruments is conserved.

\section{Application to the US term structure before the Coronavirus recession }\label{section:empirical}

The aim of the section is to analyse the behavior in both models in a real situation treating as the case study the ongoing events. The objective of the analysis is to show the advantages of the use of the ZBDT model 
in aspects of the financial crisis, mainly in pre-crisis moments.

In 2019, the international financial markets showed good returns. In particular, short-term interest rates in the United States sovereign market were greater than $2 \%$ for a long period. The expectations of the financial agents that there would be an economic crisis in 2020 were very low. These expectations are possible to observe in the evolution in some stock market indices. Among them is the VIX Index which measures volatility in the Chicago market options (CBOE or Chicago Board Options Exchange) in the S\&P $500$ Index. VIX is known as the "Investor feeling index" according to which if it is less than $20$ (like in $2019$), one assumes that the investors have confidence in the US economy. However, the COVID-19 pandemic generated instability in all financial markets. During the evolution of the pandemic, there were key moments in the behavior of the main variables in the  market. International markets reacted strongly when the first cases of infections appeared in
Europe. At the end of February 2020, it was detected that several
financial agents disposed in a massive way their positions (with greater risk)
and reinvested their proceeds into the "safe" assets. Then followed a March, during which a sharp global economic downturn was observed and the stock markets suffered heavy losses. As a result of the situation, on March 15th the Federal Reserve System embarked on a large-scale bond buying operation and introduced additional measures to support the economy. The objective of such proceedings was the reactivation of consumption and markets liquidity.

\begin{figure}[h!]
\centering \includegraphics[scale=0.45]{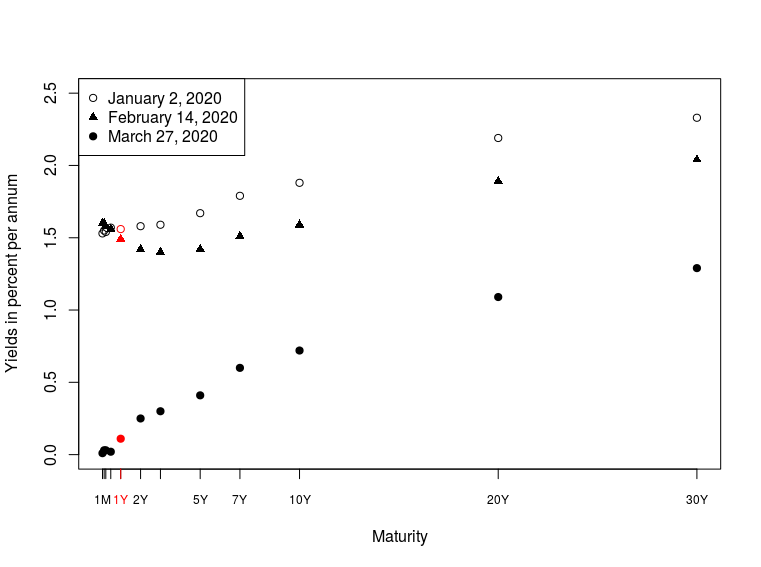}
\caption{ United States term structure for different stages of COVID-19 pandemic. The 1-year yield is marked in the red colour.}\label{term_structure}
\end{figure}

In Figure \ref{term_structure}, we present the term structure of the United States  in three different stages of the COVID-19 pandemic in 2020. At the beginning of January there were some records of COVID-19 confirmed infections (only in China), but there were no records of COVID-19 deaths. The international financial markets were not affected in this period. Second,  February 14 was the last day of the period, when Europe had no records of the Coronavirus deaths. So far the continent had had only individual cases of the confirmed infection. The third of the considered days is March 27 - around 2 weeks after  the World Health Organisation (WHO) announced COVID-19 outbreak a pandemic and when the Fed had been introducing drastic measures to diminish effects of the ongoing COVID-19 financial crisis. This day represents a time period when new cases of COVID-19 appeared on a massive scale in different developed countries, including the United States \cite{stat3, stat2, stat1}. In Figure \ref{term_structure}, we can observe the abrupt drop in the yield rates on sovereign assets of the United States after the monetary policy by the Fed was imposed. It is worth mentioning that short-term yield rates  decreased  in a  short period of time. For example, the US Treasury 1-year yield decreased from approximately $1.5 \%$ to $0.2 \%$  in approximately three to
four weeks.

\begin{table}[h]
\centering
\scalebox{1}{
\begin{tabular}{|ccccc|c|cccccc|}
\hline
        &         &         &         &          & &         &       &       &       &      &  100   \\
        &         &         &         & 6.05  & &         &       &       &       &  94.29 &  100   \\
        &         &         & 3.96 & 2.49  & &         &       &       &   92.28 & 97.57  &  100  \\
        &         & 2.73 & 1.72 & 1.03  & &         &            & 91.94&  96.62 &98.99&100 \\
        & 1.86 & 1.14 & 0.75 & 0.42  & &         & 92.49 &   96.62 &  98.55& 99.58 & 100\\
 1.49 & 0.84 & 0.48 & 0.32 & 0.17  & &93.19 & 96.68 & 98.50 &99.38 &99.83 & 100   \\
          \hline
\end{tabular}}
\caption{BDT trees for interest rate ($\%$) and zc-bond prices for the Real Case.}\label{tabla_apl_bdt}
\end{table}

\begin{table}[h]
\centering
\scalebox{1}{
\begin{tabular}{|ccccc|c|cccccc|}
\hline
        &         &         &         &          & &         &       &       &       &      &  100   \\
       &         &         &         & 5.77   & &     &       &       &       &  94.54    &  100   \\
       &         &         & 4.07 & 2.73   & &     &       &       &      92.19 &97.35&  100   \\
       &         & 3.08 & 1.93 & 1.29   & &     &       &       91.37 &96.18& 98.73 &  100   \\
       & 2.03 & 1.19 & 0.91 & 0.61   & &     &      91.84& 96.03 & 98.17 & 99.40    &  100   \\
1.49 & 0.93 & 0.66 & 0.64 & 0.30   & &   93.19 & 96.35& 98.00 &98.93 &99.70  &  100   \\
       & 0.25 & 0.25 & 0.25 & 0.25   & &     &    98.99& 99.25& 99.50& 99.75    &  100   \\
          \hline
\end{tabular}}
\caption{ZBDT trees for interest rate ($\%$) and zc-bond prices for the Real Case.}\label{tabla_apl_zbdt}
\end{table}

 We analyse the day before the high volatility in the financial markets appeared. From our perspective, the ZBDT model is a good tool for valuing financial derivatives in interest rates before the COVID-19 pandemic. This statement derives from the fact that the classic models have no probability of such an abrupt rate decrease in a short period of time. In this application (we will call it "Real Case"), we use the first five years of the term structure published in U.S. Department of the Treasury for February 14th (see the data in \cite{treasury}). These published yields (in $\%$) and the annual volatility of the rates estimated (in $\%$) can be found in equation \ref{yield-vol-caso-real} below.  For the ZBDT model, we consider the parameters $ p = 0.1 $, $ x_{0} = 0.25\% $ and $q=0.01 $. Note that the value $p=0.1$ corresponds to the higher risk of the recession (that is related to the beginning of the COVID-19 pandemic) compared to the value $p=0.02$ assumed for the more "conservative" Example in Section $4$. Table \ref{tabla_apl_bdt} and Table \ref{tabla_apl_zbdt} above contain the interest rate and the zc-bond price trees for the BDT and ZBDT models. With these results it is possible to value the selected financial  derivatives. Figure \ref{impVolsec6} below presents the option prices in dependence of the strike for BDT (red) and ZBDT model (blue) for the selected European call options: (a) Knock-Up-Out, (b) Knock-Double-Out, (c) Knock-Down-In, (d) Knock-Down-Out, (e) Knock-Up-In, (f) Knock-Double-In option. The barriers are $H^{-}=93$ and $H^{+}=98.5$.  Similarly as in the case of Figure \ref{implvolsec5}, we conclude that the relation between the curves of the option prices corresponding to BDT and ZBDT model are different depending on the option contract.
 
 \begin{align}
\label{yield-vol-caso-real}
yield &= (1.49,\ 1.42,\ 1.40,\ 1.41,\ 1.42); \nonumber\\
Vol&=(25.5,\ 39.8,\ 41.7,\ 41.6,\ 42.2).
\end{align}

\begin{figure}[h]
\begin{center}
\subfigure[]{
\includegraphics[width=0.25\textwidth]{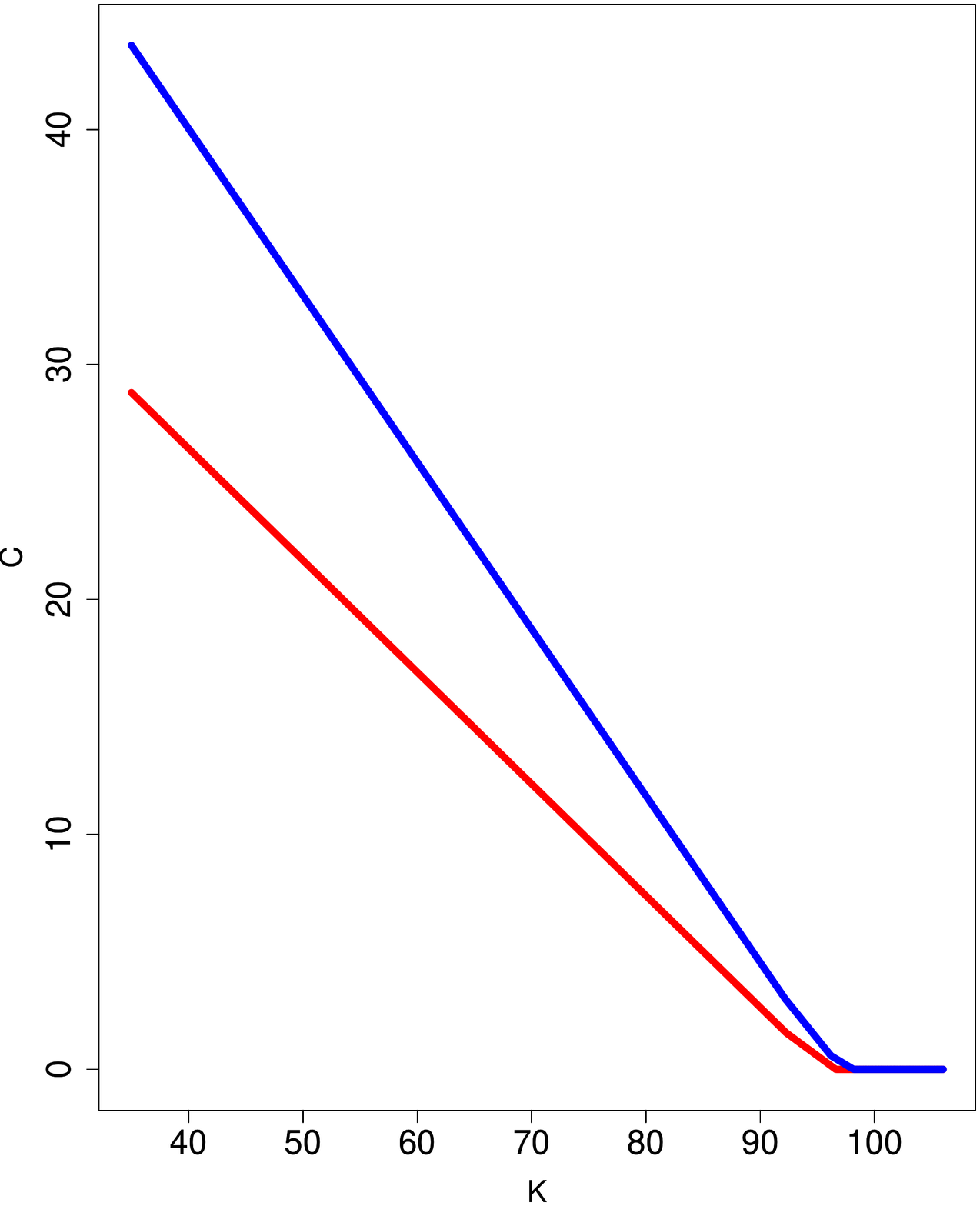}
}
\subfigure[]{
\includegraphics[width=0.25\textwidth]{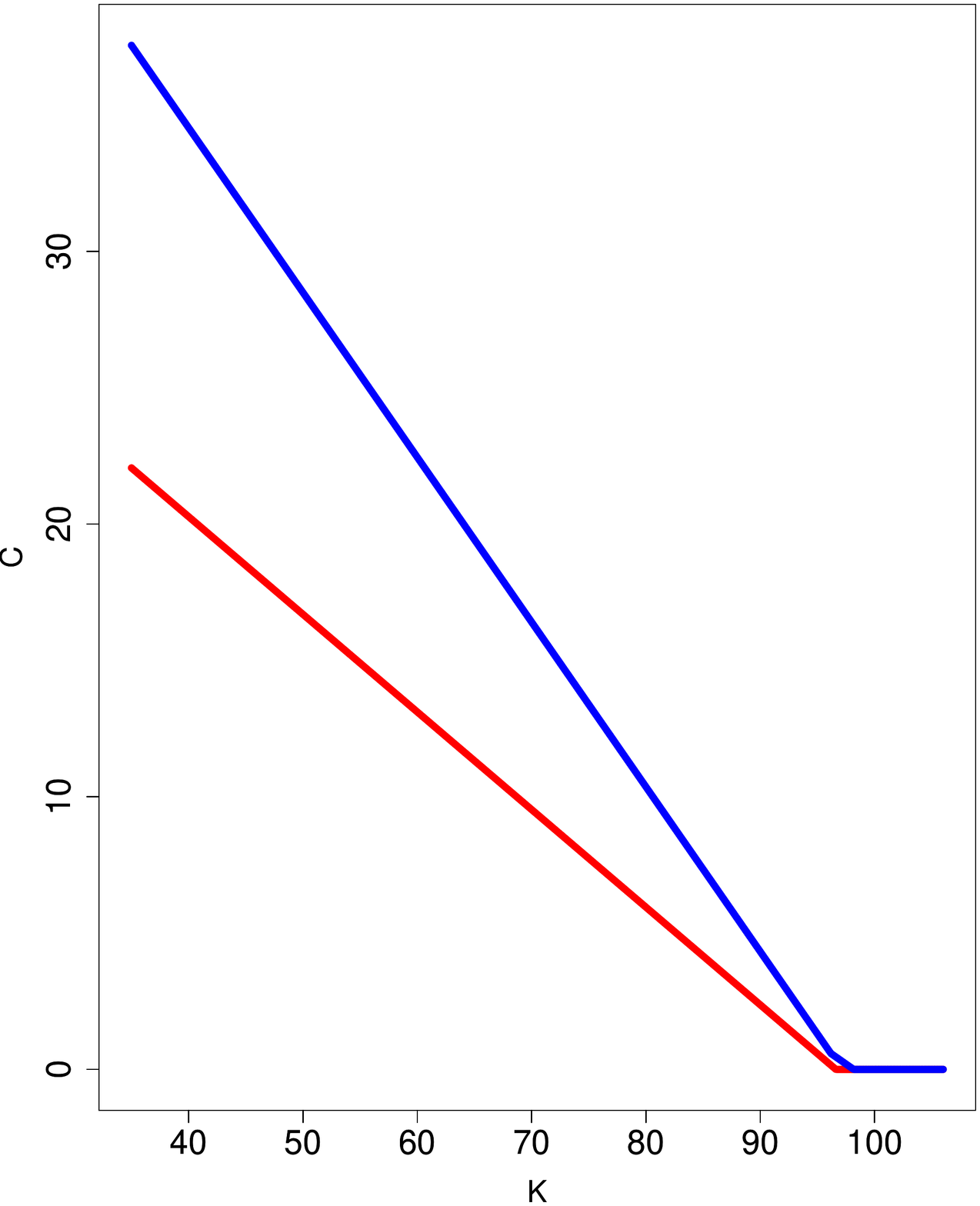}
}
\subfigure[]{
\includegraphics[width=0.25\textwidth]{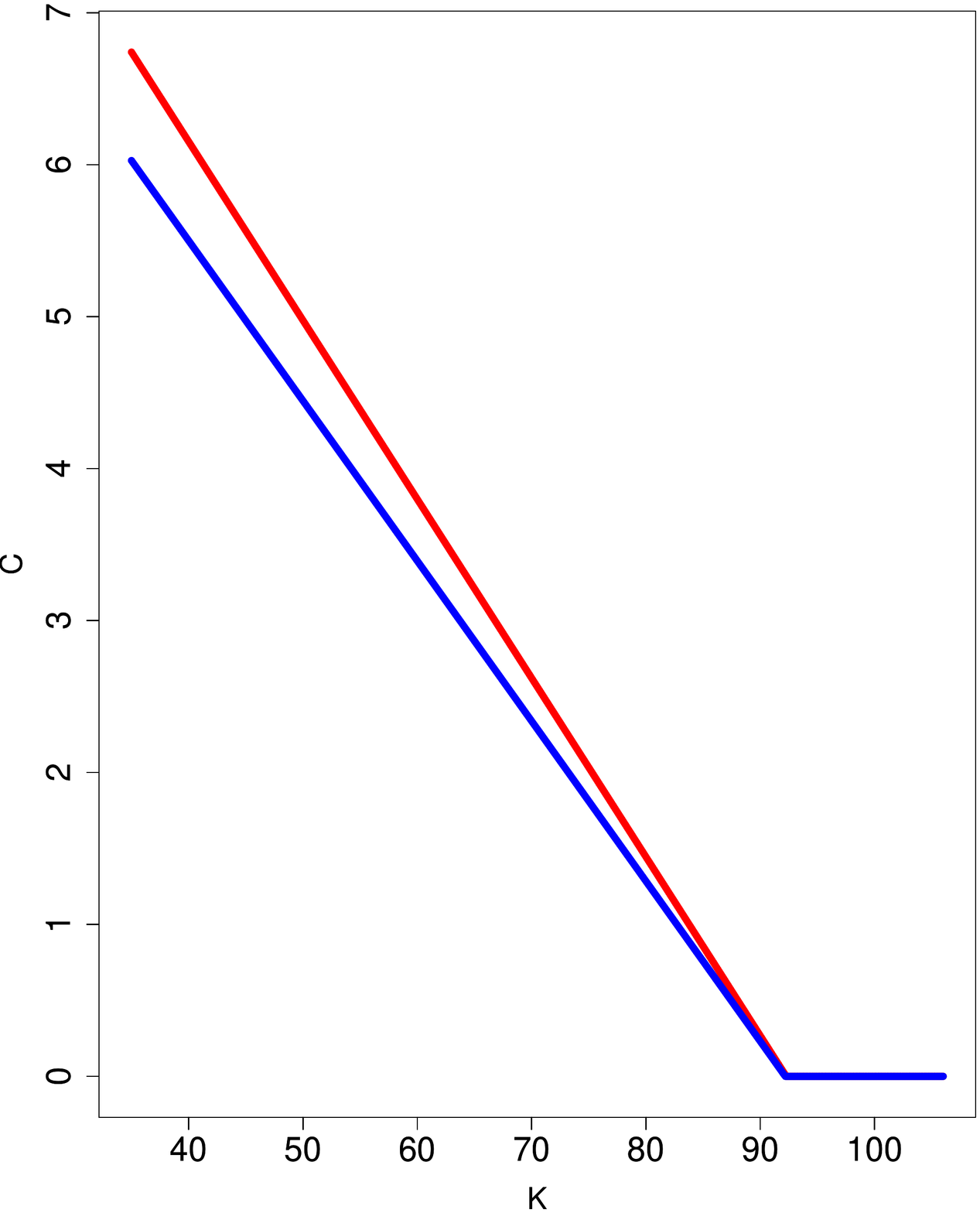}
}
\\
\subfigure[]{
\includegraphics[width=0.25\textwidth]{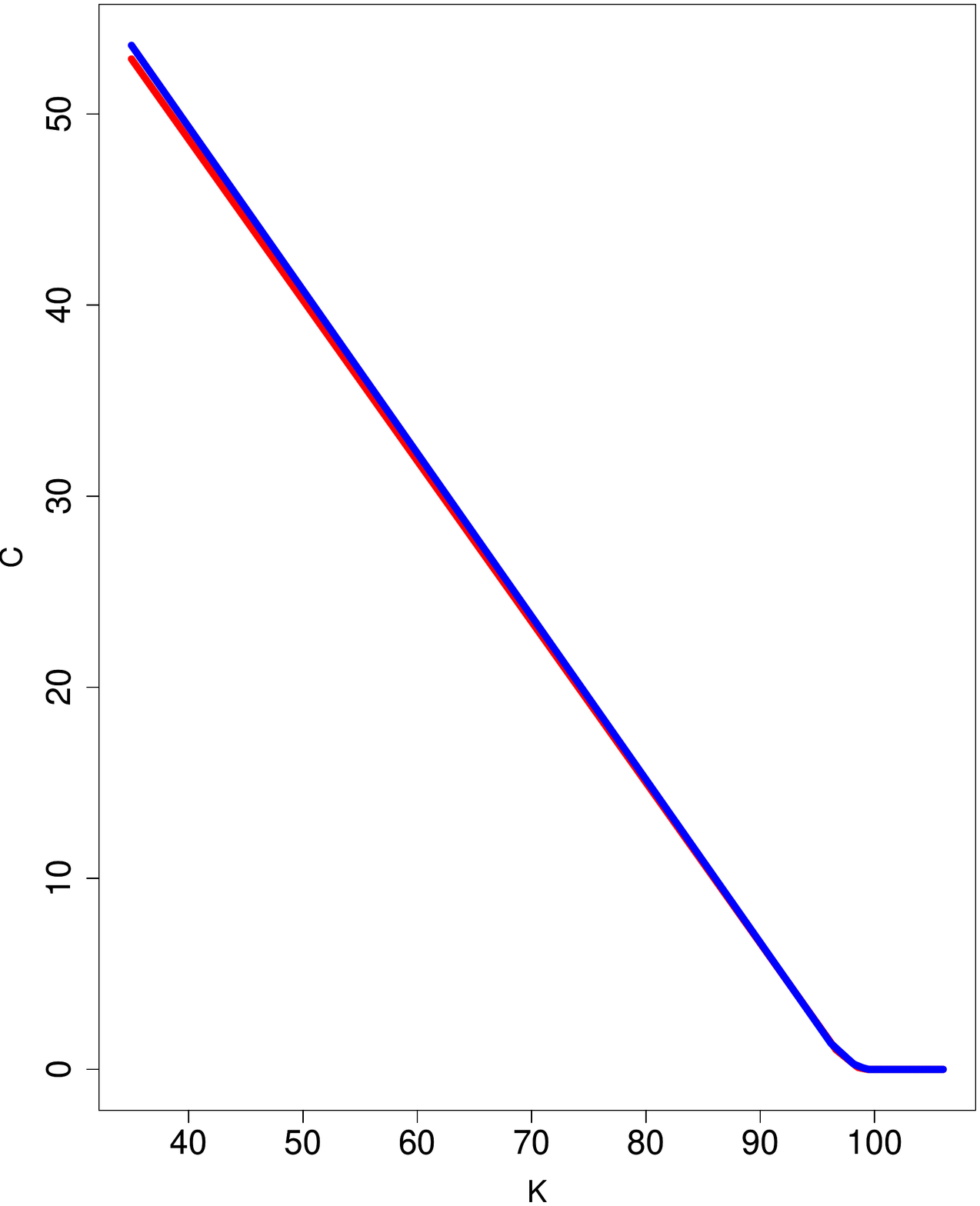}
}
\subfigure[]{
\includegraphics[width=0.25\textwidth]{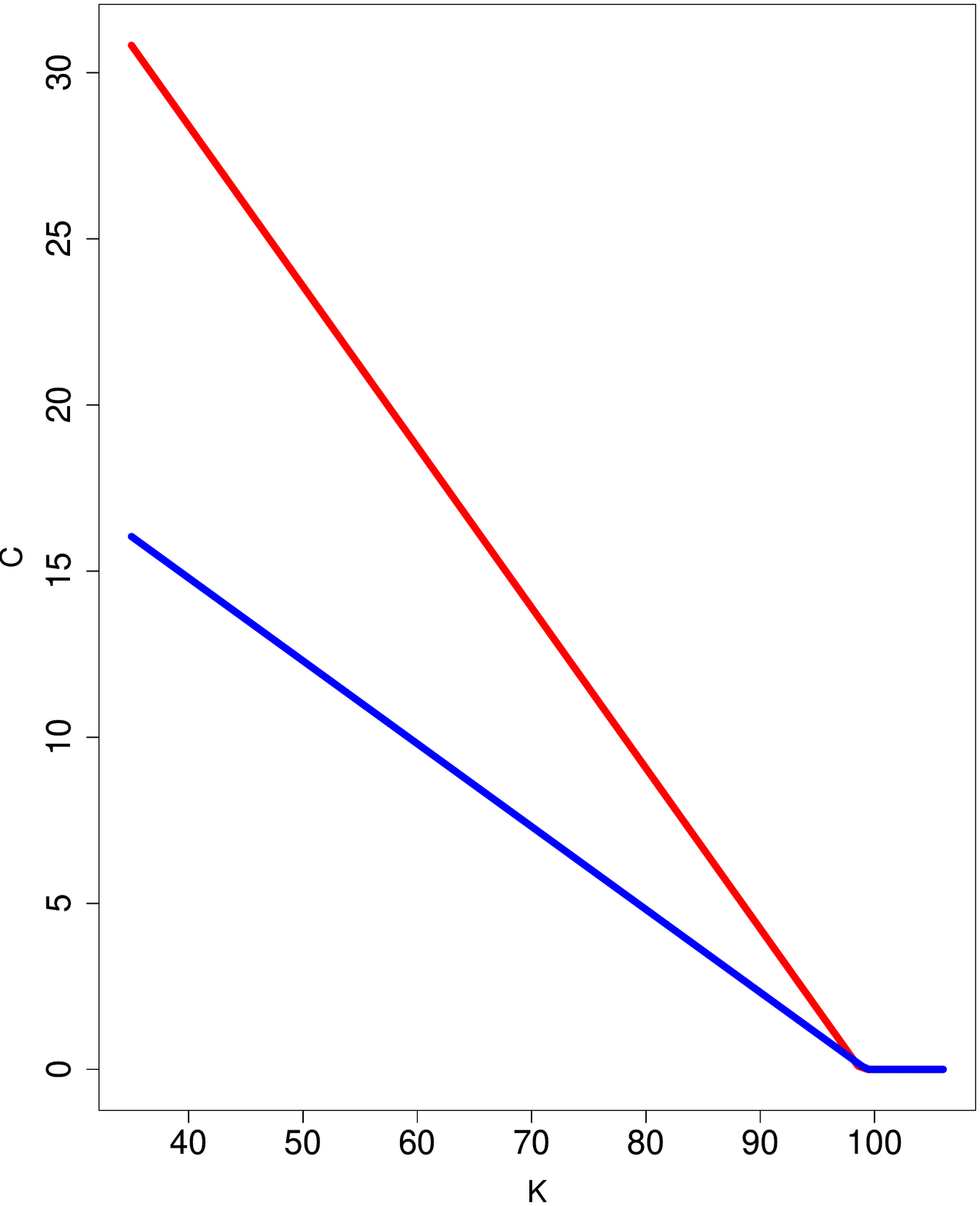}
}
\subfigure[]{
\includegraphics[width=0.25\textwidth]{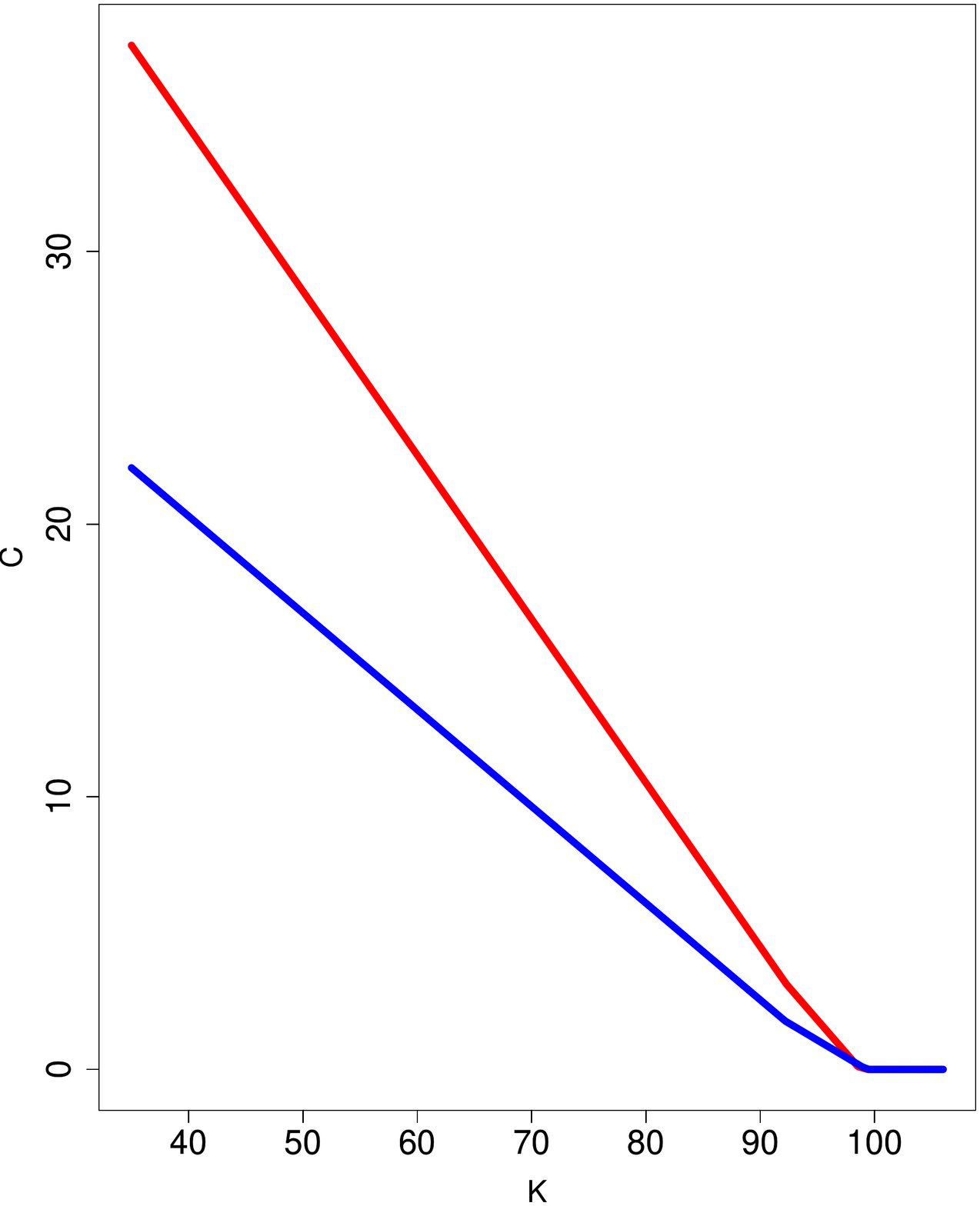}
}
\caption{Option prices in dependence of the strike price $K$ for ZBDT model (blue) and BDT model (red) for different Barrier options on bonds from Tables \ref{tabla_apl_bdt} and \ref{tabla_apl_zbdt}.  }\label{impVolsec6}
\end{center}
\end{figure}

\newpage
\section{Conclusions}\label{section:conclusions}
In this paper we propose a general methodology to price a wide range of financial derivatives as set out in the Zero Black-Derman-Toy model proposed in \cite{zbdt}.
The model  is inspired by Lewis's  ZIRP models in continuous time (see \cite{lewis}),
and also in
Duffie and Singleton's  default framework of bond pricing models (see \cite{DuffieSingleton}).
Its novelty is to add a new
branch at each period to the classical BDT tree model that includes the small probability of falling into recession or a catastrophic event, necessitating a rapid interest rate reduction to near-zero rates. The methodology for estimating interest rates $ r_ {ij} $, pricing bonds and pricing their derivatives in the ZBDT model follows the same ideas as in the BDT model.  

In a frame of the hypothetical case study, we considered a decreasing-increasing term structure and decreasing volatility. We valued different Vanilla and Barrier options in both models. Moreover we compared the implied volatility determined for European call options using Black's formula. As a result, large differences between both models are observed. We can conclude that considering the non-zero probability that interest rates will contract to values close to zero in a short period of time can match the market conditions better than the classical model.

In order to apply the model to a real US-term structure, the financial crisis generated by the Coronavirus is studied.  It is widely known that the financial markets suffered great turbulence due to the health and economic crisis caused by the  pandemic. On March 14, the  Federal Reserve System made the decision to cut the interest rate to values close to $0 \%$ and promised to buy $\$700$ billion in Treasury-backed securities and mortgages. These measures sought to prevent market disruptions from aggravating what is likely to be a severe slowdown from the pandemic. This situation leads the financial industry to improve the valuation of financial derivatives to consider catastrophic events in a more realistic way. In order to analyse both models (classical BDT and novel ZBDT) in the current situation, they were calibrated on a day before the first death caused by the Coronavirus in Europe, whose situation immediately affected the international Stock Exchanges. The BDT model and ZBDT model differ in their interest rate/bond/derivative trees. Both models can be applied for economic crises or catastrophic events but in our opinion the ZBDT model better matches the actions undertaken by Fed in the presence of the economic downturn.

   \section*{Acknowledgements}
The research of G.K. was partially supported by NCN Sonata Bis 9 grant nr 2019/34/E/ST1/00360.


 \bibliography{zbdt}{}

\begin{thebibliography}{10}

\bibitem{black}
Fischer Black.
\newblock {T}he pricing of commodity contracts.
\newblock {\em Journal of financial economics}, 3(1-2):167--179, 1976.

\bibitem{bdt}
Fischer Black, Emanuel Derman, and William Toy.
\newblock A one-factor model of interest rates and its application to treasury
  bond options.
\newblock {\em Financial analysts journal}, 46(1):33--39, 1990.

\bibitem{epid2}
Laura~SP Bloomfield, Tyler~L McIntosh, and Eric~F Lambin.
\newblock {H}abitat fragmentation, livelihood behaviors, and contact between
  people and nonhuman primates in {A}frica.
\newblock {\em Landscape Ecology}, 35(4):985--1000, 2020.

\bibitem{expl3}
Phelim~P Boyle, Ken~Seng Tan, and Weidong Tian.
\newblock {C}alibrating the {Black-Derman-Toy} model: some theoretical results.
\newblock {\em Applied Mathematical Finance}, 8(1):27--48, 2001.

\bibitem{crr}
John~C Cox, Stephen~A Ross, and Mark Rubinstein.
\newblock {O}ption pricing: {A} simplified approach.
\newblock {\em Journal of financial Economics}, 7(3):229--263, 1979.

\bibitem{DuffieSingleton}
Darrell Duffie and Kenneth~J Singleton.
\newblock {M}odeling term structures of defaultable bonds.
\newblock {\em The review of financial studies}, 12(4):687--720, 1999.

\bibitem{eberlein}
Ernst Eberlein, Christoph Gerhart, and Zorana Grbac.
\newblock {M}ultiple curve {L}{\'e}vy forward price model allowing for negative
  interest rates.
\newblock {\em Mathematical Finance}, 2018.

\bibitem{frunza2015introduction}
Marius-Cristian Frunza.
\newblock {\em {Introduction to the Theories and Varieties of Modern Crime in
  Financial Markets}}.
\newblock Academic Press, 2015.

\bibitem{ja3}
Grzegorz Krzy{\.z}anowski and Marcin Magdziarz.
\newblock {A} weighted finite difference method for {A}merican and {B}arrier
  options in subdiffusive {Black-Scholes Model}.
\newblock {\em arXiv preprint arXiv:2003.05358}, 2020.

\bibitem{zbdt}
Grzegorz Krzy{\.z}anowski, Ernesto Mordecki, and Andr{\'e}s Sosa.
\newblock {Z}ero {Black-Derman-Toy} interest rate model.
\newblock {\em arXiv preprint arXiv:1908.04401v2}, 2019.

\bibitem{lewis}
Alan~L Lewis.
\newblock {Option Valuation under Stochastic Volatility II}.
\newblock {\em Finance Press}, 2009.

\bibitem{martin}
Marcus~RW Martin.
\newblock {An Overview of Post-crisis Term Structure Models}.
\newblock In {\em New Methods in Fixed Income Modeling}, pages 85--97.
  Springer, 2018.

\bibitem{mcd}
Robert~Lynch McDonald, Mark Cassano, and R{\"u}diger Fahlenbrach.
\newblock {\em {D}erivatives markets}.
\newblock Addison-Wesley Boston, 2006.

\bibitem{epid1}
Carl-Johan Neiderud.
\newblock {H}ow urbanization affects the epidemiology of emerging infectious
  diseases.
\newblock {\em Infection ecology \& epidemiology}, 5(1):27060, 2015.

\bibitem{treasury}
U.S.~Department of~the Treasury.
\newblock {Daily} {Treasury} {Yield} {Curve} {Rates}.
\newblock
  \url{https://home.treasury.gov/policy-issues/financing-the-government/interest-rate-statistics}.
\newblock [Online; accessed 08 June 2020 3:20pm].

\bibitem{expl2}
Dan Pirjol.
\newblock {E}xplosive {Behavior} in the {Black–Derman–Toy Model}.
\newblock {\em Interdisciplinary Topics in Applied Mathematics, Modeling and
  Computational Science}, pages 361--366, 2015.

\bibitem{expl1}
Dan Pirjol.
\newblock {Hogan--Weintraub} singularity and explosive behaviour in the
  {Black--Derman--Toy} model.
\newblock {\em Quantitative Finance}, 15(7):1243--1257, 2015.

\bibitem{sifma}
SIFMA.
\newblock {SIFMA Capital Markets Fact Book}, 2019.
\newblock
  \url{https://www.sifma.org/wp-content/uploads/2019/09/2019-Capital-Markets-Fact-Book-SIFMA.pdf},
  2019.
\newblock [Online; accessed 29 April 2020 9:15pm].

\bibitem{stat3}
Statista.
\newblock {Number} of coronavirus ({COVID-19}) deaths in {E}urope since
  {F}ebruary 2020 (as of {May} 24, 2020), by country and date of report.
\newblock
  \url{https://www.statista.com/statistics/1102288/coronavirus-deaths-development-europe/}.
\newblock [Online; accessed 03 June 2020 9:25pm].

\bibitem{stat2}
Statista.
\newblock {Number} of new coronavirus ({COVID-19}) cases in {E}urope from
  {J}anuary 25 to {J}une 2, 2020, by date of report.
\newblock
  \url{https://www.statista.com/statistics/1102209/coronavirus-cases-development-europe/}.
\newblock [Online; accessed 03 June 2020 9:20pm].

\bibitem{skew}
Yingxu Tian and Haoyan Zhang.
\newblock {S}kew {CIR} process, conditional characteristic function, moments
  and bond pricing.
\newblock {\em Applied Mathematics and Computation}, 329:230--238, 2018.

\bibitem{stat1}
Worldometer.
\newblock {COVID-19} {C}oronavirus pandemic.
\newblock \url{https://www.worldometers.info/coronavirus/}.
\newblock [Online; accessed 03 June 2020 9:15pm].

\end{thebibliography}

\bibliographystyle{plain}

\end{document}